\documentclass[aps,prd,twocolumn,showpacs,nofootinbib,eqsecnum,floatfix]{revtex4}

\usepackage{amsmath,amssymb}
\usepackage{graphicx}

\newcommand{\half}{{\textstyle\frac{1}{2}}}

\newcommand{\threehalf}{{\textstyle\frac{3}{2}}}

\newcommand{\fourth}{{\textstyle\frac{1}{4}}}

\newcommand{\Tr}{{\rm Tr\hskip 2pt}}
\def\lsim{\mathrel{\rlap{\raise 2.5pt \hbox{$<$}}\lower 2.5pt\hbox{$\sim$}}}
\def\gsim{\mathrel{\rlap{\raise 2.5pt \hbox{$>$}}\lower 2.5pt\hbox{$\sim$}}}

\def\vecalpha{{\pmb\alpha}}

\renewcommand{\Re}{{\rm Re\thinspace}}
\renewcommand{\Im}{{\rm Im\thinspace}}

\allowdisplaybreaks 

\begin{document}

\title{Trilinear Higgs couplings in the two Higgs doublet model 
        with CP violation}

%
\author{Per Osland}
\email[]{per.osland@ift.uib.no}
\affiliation{Department of Physics and Technology, University of Bergen,
Postboks 7803, N-5020 Bergen, Norway}
\author{P. N. Pandita}
\email[ ]{ppandita@nehu.ac.in}
\affiliation{Service de Physique Th\'eorique, CEA-Saclay,
F-91191 Gif-sur-Yvette Cedex, France and\\
Department of Physics, North Eastern Hill University,
Shillong 793 022, India\footnote{Permanent address}}
\author{Levent Selbuz}
\email[]{Levent.Selbuz@eng.ankara.edu.tr}
\affiliation{Department of Physics and Technology, University of Bergen,
Postboks 7803, N-5020 Bergen, Norway and \\
Department of Engineering Physics, Faculty of
Engineering, Ankara University, 06100 Tandogan-Ankara, 
Turkey\footnote{Permanent address}}

\date{\today}

\begin{abstract}
We carry out a detailed analysis of the general two Higgs doublet model with CP
violation. We describe two different parametrizations of this model,
and then study the Higgs boson masses and the trilinear Higgs couplings 
for these two parametrizations. Within a rather general model, we find that the
trilinear Higgs couplings have a significant dependence on the details of the
model, even when the lightest Higgs boson mass is taken to be a fixed
parameter. We include radiative corrections in the one-loop effective
potential approximation in our analysis of the Higgs boson masses and the
Higgs trilinear couplings. The one-loop corrections to the trilinear couplings
of the two Higgs doublet model also depend significantly on the details of the
model, and can be rather large.  We study quantitatively the trilinear Higgs
couplings, and show that these couplings are typically several times larger
than the corresponding Standard Model trilinear Higgs coupling in some 
regions of the parameter space.  We also
briefly discuss the decoupling limit of the two Higgs doublet model.
\end{abstract}

\pacs{12.60.Fr, 14.80.Cp, 11.30.Er}
\keywords{Two Higgs doublet model, trilinear couplings, CP violation}

\maketitle

\section{Introduction}
\label{sec:introduction}
The Higgs mechanism~\cite{Englert:1964et} of spontaneous electroweak symmetry
breaking is a necessary ingredient of the Standard Model~(SM), which is
crucial for its internal consistency.  The search for the Higgs boson is,
thus, one of the major tasks for the experiments at the upcoming Large Hadron
Collider~(LHC). In order to confirm the Higgs mechanism as the origin of
spontaneous breaking of $SU(2)_L \times U(1)_Y$ gauge symmetry in the SM, not
only must the Higgs boson be discovered, but also its
trilinear~($\lambda_{HHH}^{\text{SM}}$) and
quartic~($\lambda_{HHHH}^{\text{SM}}$) self couplings must be measured in
order to completely reconstruct the Higgs potential.  Furthermore, one must
also be able to measure the couplings of the Higgs boson to gauge bosons and
fermions.

Here we shall be concerned mainly with the trilinear self couplings of the
Higgs boson. In the Standard Model, there is only one trilinear self coupling
of the Higgs boson which can be written simply in terms of the Higgs boson
mass $M_H$ as
\begin{equation} \label{Eq:SM-coupling}
\lambda_{HHH}^{\text{SM}}=\frac{3M_H^2}{v},
\end{equation}
where $v = 2M_W/g = 246 $~GeV is the vacuum expectation value of the neutral
component of the Higgs doublet, $M_W$ is the mass of the $W^{\pm}$, and $g$ is
the $SU(2)_L$ gauge coupling, respectively. Several extensions of the SM, such
as the minimal supersymmetric standard model~(MSSM)~\cite{Nilles:1983ge}, and
the general two Higgs doublet model~(2HDM)~\cite{Gunion:1989we}, have a more
complicated Higgs structure. In these models there are several trilinear Higgs
couplings, having more complicated dependence on the underlying masses. It is
a challenging task to measure~\cite{Weiglein:2004hn} these trilinear couplings
at the LHC.  On the other hand a linear collider could possibly offer much
better prospects of measuring these trilinear Higgs 
couplings~\cite{Accomando:1997wt,Djouadi:1996ah,AguilarSaavedra:2001rg,
Arhrib:2008jp,Ferrera:2007sp}.

At the proposed International Linear Collider~(ILC), if the Higgs boson is not
too heavy, the trilinear Higgs coupling can be measured via the Higgs boson 
pair production 
$e^+e^-\rightarrow W^{+*}\overline{\nu}W^{-*}\nu \to HH\overline{\nu}\nu$.  
A precise measurement of the trilinear Higgs self coupling will also make 
it possible to test extended Higgs models, which have a different structure 
of the Higgs potential, and hence different trilinear Higgs couplings, as 
compared to the SM. The 2HDM is the simplest, yet a very general, model with 
an extended Higgs structure, which leads to various distinct physical effects. 
In particular, the model can easily accommodate additional CP violation
~\cite{Lee:1973iz,Weinberg:1976hu,Branco:1985aq,Accomando:2006ga}, beyond what 
is generated by the Kobayashi--Maskawa~(KM) mechanism~\cite{Cabibbo:yz}.
Furthermore, since there is a distinct possibility of measuring precisely 
the Higgs boson self couplings at the ILC, there is a motivation to study 
the  radiative corrections to the trilinear self couplings of the Higgs boson. 

On the other hand,  there are a number of parameters in the
potential of the two Higgs doublet model, including those associated with 
CP violation, which determine the
masses and CP properties of the model. In spite of the complicated nature of
the Higgs potential, it has been shown in the 2HDM with CP conservation that
one-loop corrections to the lightest CP-even Higgs boson self coupling are 
in general of
a non decoupling nature, and can give rise to $\mathcal{O}(100\%)$ deviations
from the SM prediction~\cite{Kanemura:2002vm}. This happens even when all
other couplings of the lightest Higgs boson to gauge bosons and fermions are
in good agreement with the SM prediction.

In this paper we shall study in detail the trilinear Higgs couplings in 
the two Higgs Doublet Model~(2HDM) with CP violation. We shall assume
that the underlying gauge group is the SM gauge group. After spontaneous 
breaking of the SM gauge symmetry, the Higgs spectrum of the model consists 
of three neutral Higgs bosons and two charged Higgs bosons. In the 
CP-conserving version of the model, two neutral Higgs bosons~($h^0, H^0$) 
are CP even, whereas one neutral Higgs boson~($A^0$) is CP odd. In the 
CP-conserving case, there are, thus,  six allowed trilinear Higgs couplings 
which can be labelled as $\lambda_{hhh}$,
$\lambda_{hhH}$, $\lambda_{hHH}$, $\lambda_{HHH}$ involving the CP-even Higgs
bosons, and $\lambda_{hAA}$, $\lambda_{HAA}$, all even in the number of $A^0$,
involving the CP-odd Higgs boson.  When CP is not conserved, the neutral Higgs
bosons do not have a definite CP, and there is no such constraint on the
couplings involving the $A^0$ Higgs boson. There, are, thus, a total of ten
trilinear Higgs couplings in the case of the 2HDM with CP violation. 

Since this is one of the simplest models which goes beyond the KM  mechanism
of CP violation, it is important to study the impact of CP violation 
on the Higgs self couplings in this model. We shall, therefore, consider
the 2HDM with explicit CP violation in the Higgs sector in this paper. 
This explicit CP violation is introduced through appropriate complex 
parameters in the potential of the 2HDM. Furthermore, we shall assume 
that this explicit CP violation cannot be transformed away by a 
redefinition of the Higgs fields.
Thus, one of the objectives of the present study is to determine the effects 
of this explicit CP violation in the Higgs sector on various Higgs self 
couplings in the two Higgs doublet model. In our study of the model,
we shall find it convenient to keep the masses of the two lightest neutral
Higgs bosons fixed, and then study the dependence on other parameters
of the model, which will determine the amount of explicit 
CP violation in various 
Higgs self couplings, and also determine the mass of the heaviest Higgs boson.
In this way, we do not lay emphasis on the heavy Higgs sector of the theory,
whereas at the same time we can exhibit the wide range of values that the 
trilinear Higgs couplings can assume. In particular, we wish to emphasize
that, in contrast to the MSSM, the heavy mass effects in a general 2HDM
do not decouple. In order to demonstrate this we shall calculate the one-loop
corrected Higgs boson self-couplings using the method of effective potential  
in a general 2HDM with explicit CP violation as described above.
This is in contrast to Ref.~\cite{Kanemura:2002vm}, where all couplings 
and mass parameters 
were assumed to be real, thereby precluding the phenomena of explicit
CP violation in the Higgs sector as considered in this paper.

The plan of this paper is as follows. In Section \ref{sec:2HDM} we describe
the most general two Higgs doublet model with CP violation. We then discuss
the spectrum of the Higgs bosons of the model, and the constraints on the
parameters when there is CP violation in the model.  Here we also discuss how
only certain sets of parameters of the Higgs potential can lead to physically
consistent models, and discuss different ways in which these parameters can be
specified. We delineate the regions of the parameter space where there is no
CP violation, corresponding to which one of the neutral Higgs bosons is odd
under P.  In Section~\ref{sect:tree-level-couplings} we derive the tree level
trilinear couplings between the neutral as well as the charged Higgs bosons of
the model, and discuss the correspondence of these couplings with the
trilinear Higgs couplings of the minimal supersymmetric standard model. Here
we also study numerically the trilinear Higgs couplings, and discuss the
domain of the parameter space that is compatible with various theoretical and
experimental constraints.  

In Section~\ref{sect:one-loop-level-masses} we use the method of one-loop
effective potential to calculate, as a first step, one-loop corrections to the
Higgs boson masses.  In Section~\ref{sect:one-loop-level-couplings}, we then
calculate the one-loop corrections to the trilinear self couplings of the
Higgs bosons.  In Section~\ref{sec:quantitative results} we carry out a
detailed numerical study of the one-loop corrected trilinear Higgs couplings,
discuss the magnitudes of different contributions, the dependence on the
scale, as well as the decoupling limit, and show that these couplings can be
several times larger than the corresponding Standard Model trilinear Higgs
coupling in some regions of the parameter space.  
We summarize our results and conclusions in
Section~\ref{sect:Summary}.  Some of the analytical calculations used in our
analysis are described in appendices.

\section{The Two-Higgs-Doublet Model}
\label{sec:2HDM}
The general 2HDM with the underlying gauge group $SU(2)_L \times U(1)_Y$
is obtained by extending  the Higgs sector of the SM with a second 
$SU(2)_L$ Higgs doublet  with weak hypercharge $Y = 1.$ Thus, the model 
contains $4$ complex scalar fields which are  arranged as $SU(2)_L$ doublets 
as follows:
\begin{equation}
 \Phi_{i}
 = \begin{pmatrix}
 \varphi_i^+\\
  \varphi_i^0
 \end{pmatrix} \quad (Y = +1), \quad i=1,\, 2.
\label{hdoublets}
\end{equation}
Using these two Higgs doublets, the most general renormalizable  potential 
for the 2HDM which is invariant under the $SU(2)_L \times U(1)_Y$ gauge 
group can be written as
\begin{align}
V_\text{tree}&=\frac{\lambda_1}{2}(\Phi_1^\dagger\Phi_1)^2
+\frac{\lambda_2}{2}(\Phi_2^\dagger\Phi_2)^2
+\lambda_3(\Phi_1^\dagger\Phi_1) (\Phi_2^\dagger\Phi_2) \nonumber \\
&+\lambda_4(\Phi_1^\dagger\Phi_2) (\Phi_2^\dagger\Phi_1) 
+\frac{1}{2}\left[\lambda_5(\Phi_1^\dagger\Phi_2)^2+{\rm h.c.}\right]
\nonumber \\
&+\left\{\left[\lambda_6(\Phi_1^\dagger\Phi_1)+\lambda_7
(\Phi_2^\dagger\Phi_2)\right](\Phi_1^\dagger\Phi_2)
+{\rm h.c.}\right\} \nonumber \\
&-\frac{1}{2}\bigl\{m_{11}^2(\Phi_1^\dagger\Phi_1)
+\left[m_{12}^2 (\Phi_1^\dagger\Phi_2)+{\rm h.c.}\right]
\nonumber \\
&+m_{22}^2(\Phi_2^\dagger\Phi_2)\bigr\},
\label{hpotential}
\end{align}
where $\lambda_i~(i = 1,\ldots,7)$ are dimensionless parameters and the
subscript ``tree'' denotes that (\ref{hpotential}) is a tree-level
potential. We note that $\lambda_i~(i = 1,\ldots,4)$ are real, whereas
$\lambda_i~(i= 5,\ldots,7)$ are in general complex parameters.  Similarly, 
$m^2_{11}$ and $m^2_{22}$ are real, whereas $m^2_{12}$ is in general complex. 
We note that the terms proportional to $\lambda_6$ and $\lambda_7$ have to be
constrained, since this potential does not satisfy natural flavor
conservation~\cite{Glashow:1976nt}, even when each doublet is coupled only to
up-type or only to down-type quarks.

When the neutral components of the two Higgs doublets $\Phi_{1,2}$ acquire
vacuum expectation values~(VEVs), $\varphi^0_{1,2} = v_{1,2}/\sqrt {2}$, the
gauge group $SU(2)_L \times U(1)_Y$ breaks down to $U(1)_{\rm em},$ whereby
three of the eight real fields in (\ref{hdoublets}) are absorbed by three of
the four gauge bosons of $SU(2)_L \times U(1)_Y,$ which become massive in the
process, leaving behind a massless photon.  We can then parametrize the two
Higgs doublet fields in (\ref{hdoublets}) as
\begin{equation}
 \Phi_{i}
 = \begin{pmatrix}
 \varphi_i^+\\
\frac{1}{\sqrt{2}}(v_i+\eta_i+i\chi_i)
 \end{pmatrix}, \quad i=1,\, 2,
\label{parhiggs}
\end{equation}
where we have chosen the VEVs of the neutral Higgs fields to be real, and 
absorbed the relative phase between the two VEVs in the parameters 
$m^2_{12}, \lambda_5, \lambda_6$ and $\lambda_7$ of $V_{\rm tree}$. 
When we substitute the parametrisation 
(\ref{parhiggs}) of the Higgs fields in the potential (\ref{hpotential}),
there will be cubic terms in the  Higgs fields 
arising from the quartic couplings $\lambda_i$, which will give rise to
trilinear couplings among the Higgs fields. It is these trilinear
couplings that we  shall study in detail in this paper.

As discussed in the Introduction,  in the case of 2HDM with CP conservation,
after spontaneous breakdown of the gauge symmetry, we are left with two
CP-even Higgs bosons $h^0, H^0,$ a CP-odd Higgs boson $A^0,$ and a pair of
charged Higgs bosons $H^{\pm}$. However, with CP violation the neutral Higgs
bosons $h^0, H^0, A^0$ mix, and it is then more appropriate to define the weak
states
\begin{equation} \label{Eq:etas}
\begin{pmatrix}
\eta_1 \\ \eta_2 \\ \eta_3
\end{pmatrix},
\end{equation}
to describe the neutral Higgs sector of the CP violating two Higgs 
doublet model. In (\ref{Eq:etas}) we have defined
\begin{align}
\eta_3 &=-\sin\beta\,\chi_1+\cos\beta\,\chi_2, \label{def_eta3} \\
G^0 &=  \cos\beta\,\chi_1+\sin\beta\,\chi_2, \label{def_g0}
\end{align}
where $G^0$ is the would-be neutral Goldstone boson,
and $\tan\beta = v_2/v_1$ is the ratio
of the vacuum expectation values of the two Higgs fields.
We shall take $\tan\beta$ as an independent parameter.
The charged Higgs fields and the corresponding charged 
Goldstone boson are likewise defined as
\begin{align}
H^{\pm} &=  - \sin\beta\, \varphi^{\pm}_1  +  \cos\beta\, \varphi^{\pm}_2, 
\label{charged_hig}\\
G^{\pm} & =  \cos\beta\, \varphi^{\pm}_1  + \sin\beta\, \varphi^{\pm}_2,
\end{align}
respectively. 

\subsection{Basis rotation}

The tree-level mass squared matrix of the neutral Higgs bosons 
can now be defined as
\begin{equation} \label{Eq:M_sq-def}
{\cal M}^2_{ij}
=\frac{\partial^2 V}{\partial\eta_i\partial\eta_j},
\end{equation}
where, after differentiation, all fields are set equal to zero:
$\eta_1=\eta_2=\eta_3=H^\pm=G^0=G^\pm=0$.  The physical neutral Higgs states
$H_i$, which are the eigenstates of the mass squared matrix
(\ref{Eq:M_sq-def}), are then obtained by a rotation $R$
\begin{equation} \label{Eq:R-def}
H=R\eta, \quad \eta=R^{\text{T}}H,
\end{equation}
or, more explicitly
\begin{equation}
H_i=R_{ij}\eta_j, \quad \eta_j=R_{ij}H_i.
\end{equation}
The rotation matrix $R$ which diagonalizes~(\ref{Eq:M_sq-def}), 
\begin{equation}
\label{Eq:cal-M}
R{\cal M}^2R^{\rm T}={\cal M}^2_{\rm diag}={\rm diag}(M_1^2,M_2^2,M_3^2),
\end{equation}
with $M_1<M_2<M_3$,
can be parametrized as 
\begin{widetext}
\begin{align}     \label{Eq:R-angles}
R=R_3\,R_2\,R_1
=&\begin{pmatrix}
1         &    0         &    0 \\
0 &  \cos\alpha_3 & \sin\alpha_3 \\
0 & -\sin\alpha_3 & \cos\alpha_3
\end{pmatrix}
\begin{pmatrix}
\cos\alpha_2 & 0 & \sin\alpha_2 \\
0         &       1         & 0 \\
-\sin\alpha_2 & 0 & \cos\alpha_2
\end{pmatrix}
\begin{pmatrix}
\cos\alpha_1 & \sin\alpha_1 & 0 \\
-\sin\alpha_1 & \cos\alpha_1 & 0 \\
0         &       0         & 1
\end{pmatrix}  \nonumber \\
=&\begin{pmatrix}
c_1\,c_2 & s_1\,c_2 & s_2 \\
- (c_1\,s_2\,s_3 + s_1\,c_3) 
& c_1\,c_3 - s_1\,s_2\,s_3 & c_2\,s_3 \\
- c_1\,s_2\,c_3 + s_1\,s_3 
& - (c_1\,s_3 + s_1\,s_2\,c_3) & c_2\,c_3
\end{pmatrix},
\end{align}
\end{widetext}
where $c_i=\cos\alpha_i$, $s_i=\sin\alpha_i$. 
Note that Eq.~(\ref{Eq:cal-M}) can be inverted as
\begin{equation}
\label{Eq:calM-RMsqR}
({\cal M}^2)_{ij}=\sum_k R_{ki} M_k^2 R_{kj}.
\end{equation}
By symmetry, $({\cal M}^2)_{ji}=({\cal M}^2)_{ij}$,
and, in the general case, this matrix is seen to contain 
6 independent parameters.
These may be taken as the three masses, and the angles
of the rotation matrix.

As discussed in Ref.~\cite{ElKaffas:2007rq}, there are three limits in which
there is no CP violation in the 2HDM, corresponding to which one of the 
neutral Higgs bosons is odd under P.  These limits can be characterised in 
terms of the angles $\alpha_2$ and $\alpha_3$ of the rotation matrix 
(\ref{Eq:R-angles}) as follows:
\begin{alignat}{2}
&H_1\text{ is odd:}&\quad &\alpha_2=\pm\pi/2, \label{Eq:H_1-odd}\\
&H_2\text{ is odd:}&\quad &\alpha_3=\pm\pi/2, \\
&H_3\text{ is odd:}&\quad &\alpha_2=0,\ \ \alpha_3=0.
\label{Eq:CPcons-3}
\end{alignat}
We note that in the latter limit, we have $\alpha_1=\alpha+\pi/2$, where
$\alpha$ is the conventional mixing angle in the CP-even sector of the MSSM
\cite{Gunion:1989we}.  Any CP violation in the 2HDM will also depend on the
Yukawa couplings.  In the 2HDM~II (Model II), where the down-type quarks
couple to $\Phi_1$ and the up-type quarks couple to $\Phi_2$, various measures
of ``maximal'' CP violation both in the Higgs-vector boson and Higgs-quark
sectors are discussed in \cite{Khater:2003ym}.  These maxima occur in the
``bulk'' of the $\alpha_2$--$\alpha_3$ space, typically for $|\alpha_2|={\cal
O}(\pi/4)$ and $|\alpha_3|={\cal O}(\pi/4)$.

\subsection{Neutral Higgs boson  masses}
\label{sect:tree-level-neut-masses}
Minimizing the tree-level potential according to
\begin{equation} \label{Eq:min-cond-tree}
\frac{\partial V}{\partial \Phi_i}=0, \quad i=1,\, 2,
\end{equation}
and using the resulting  minimization conditions
to eliminate  $m_{11}^2$ and $m_{22}^2,$ one obtains the 
elements (\ref{Eq:M_sq-def}) of the tree-level mass squared matrix 
\begin{align}
\label{Eq:M^2_ij}
{\cal M}^2_{11}&=v_1^2\,\lambda_1 +v_2^2\,\nu
+\frac{v_2}{2v_1}\Re(3v_1^2\,\lambda_6-v_2^2\,\lambda_7),
\nonumber \\
{\cal M}^2_{22}&=v_2^2\,\lambda_2+v_1^2\,\nu
+\frac{v_1}{2v_2}\Re(-v_1^2\,\lambda_6+3v_2^2\,\lambda_7),
\nonumber \\
{\cal M}^2_{33}&=v^2\Re[-\lambda_5+\nu-\frac{1}{2v_1 v_2}
(v_1^2\,\lambda_6 +v_2^2\,\lambda_7)], \nonumber \\
{\cal M}^2_{12}&=v_1 v_2[\Re(\lambda_3+\lambda_4+\lambda_5)-\nu] \nonumber \\
&+\threehalf\Re(v_1^2\,\lambda_6+v_2^2\,\lambda_7), \nonumber \\
{\cal M}^2_{13}&=-\half v\, \Im[v_2\,\lambda_5+2v_1\,\lambda_6], \nonumber \\
{\cal M}^2_{23}&=-\half v\, \Im[v_1\,\lambda_5+2v_2\,\lambda_7],
\end{align}
where we have defined
\begin{align}      \label{Eq:vevs-tanbeta}
v_1 & =  v\cos\beta,\quad
v_2 = v\sin\beta,\quad \beta \in \left(0,\,\frac{\pi}{2}\right), 
\end{align}
and
\begin{equation}  \label{Eq:define-mu-nu}
\nu=\frac{1}{2v_1v_2}\,\Re(m_{12}^2),
\end{equation}
with $v^2   = v_1^2 + v_2^2 = (246\, {\rm GeV})^2.$ 
The squared masses $M_i^2$ of the physical neutral Higgs bosons are
obtained as the eigenvalues of the mass squared matrix~(\ref{Eq:M_sq-def}).
These eigenvalues are  solutions of a cubic equation which 
involves the parameters  $\lambda_i$. However, only some set of values of
the parameters $\lambda_i$ lead to consistent solutions.  In order to 
identify physically consistent  models, we shall, following the approach
of Refs.~\cite{Khater:2003wq,ElKaffas:2006nt,WahabElKaffas:2007xd}, 
specify physical~(tree-level) masses, instead of
the $\lambda_i$, as parameters.
Depending on how much constrained a model we want to consider,
we shall do this in two ways, which we shall denote ``approach (A)'' and 
``approach (B)'': 
\begin{itemize}
\item[(A)] In this case, which corresponds to $\lambda_6=\lambda_7=0$, two
elements of ${\cal M}^2$ are related via $\tan\beta$ and we cannot therefore
take all three masses as independent.  Instead, we take the {\it two
lightest} neutral Higgs boson masses, together with $\tan\beta$ and the three
angles ($\alpha_1,\alpha_2,\alpha_3$) defining the mixing matrix $R$ of
Eq.~(\ref{Eq:R-def}), as independent parameters. The third~(heaviest) Higgs
boson mass can then readily be determined.  In general, the elements $R_{13}$
and $R_{23}$ of the rotation matrix $R$ must be non-zero in order to have CP
violation. For consistency, the derived quantities $\Im\lambda_5$ and $\Im
m_{12}^2$ must be non-zero.
\item[(B)]  Here we take $\lambda_6$,  $\lambda_7 \ne 0.$  In this case, 
we take all {\it three} neutral Higgs boson masses, 
$\tan\beta$, together with the mixing matrix $R$ of Eq.~(\ref{Eq:R-def}),
$\Im\lambda_5$, $\Re\lambda_6$ and $\Re\lambda_7$ as the input parameters.
\end{itemize}
In either case, specifying the charged Higgs boson mass, $M_{H^\pm}$, as well
as the bilinear parameter $\mu^2=v^2\nu$, we can determine the remaining
$\lambda_i$ through a set of linear relations. They are, thus, unique.  
Details are given in Appendix~\ref{App:quartic}.

We note that in general, there exist multiple minima in the 2HDM potential
\cite{Lee:1973iz,Barroso:2007rr}.  However, with our choice of input
parameters, including Higgs squared masses, and these being positive, the
minimum we are working in is a global one and hence stable
\cite{Barbieri:2006dq}.  

\section{Trilinear Self Couplings of Higgs Bosons at the Tree Level}
\label{sect:tree-level-couplings}

In this Section we shall define the trilinear Higgs self couplings, and obtain
explicit expressions for them in terms of the parameters of the tree-level
Higgs potential of the 2HDM. We shall then study the behavior of these
couplings as functions of the various parameters of the model.
\subsection{Trilinear Couplings of Neutral Higgs Bosons}
\label{subsect:tree-level-couplings-neutral}
The trilinear self couplings of the neutral Higgs bosons are defined as
\begin{equation}
\lambda_{ijk}
=\frac{-i\,\partial^3 V}{\partial H_i\partial H_j\partial H_k},
\label{eq:tree-level-1}
\end{equation}
which are most easily obtained from the corresponding derivatives of $V$ in
(\ref{hpotential}) with respect to the weak fields $\eta_\ell$.  The
derivatives with respect to $H_i$ in (\ref{eq:tree-level-1}) are obtained by
noting the useful relation
\begin{equation}\label{Eq:h-R-eta}
\frac{\partial}{\partial H_i}
=\frac{d\eta_j}{d H_i}\,\frac{\partial}{\partial\eta_j}
=R_{ij}\,\frac{\partial}{\partial\eta_j},
\end{equation}
which can be used to  go from the $\eta_i$ basis to the physical $H_i$ basis.

When there is  $CP$ violation, the trilinear couplings among the physical
Higgs bosons will involve elements of the rotation matrix $R$
\cite{Choi:1999uk,Carena:2002bb}. We can then write the trilinear couplings in
terms of the derivatives of the potential (\ref{hpotential}) with
respect to $\eta_\ell$ and the elements of the rotation matrix $R$ as
\begin{align}  \label{Eq:neutral-3-coupl}
\lambda_{ijk}
&=\sum_{m\le n\le o=1,2,3}^\ast R_{i'm}R_{j'n}R_{k'o}\,
\frac{-i\,\partial^3 V}{\partial \eta_m\partial \eta_n \partial \eta_o} 
\nonumber \\
&=\sum_{m\le n\le o=1,2,3}^\ast R_{i'm}R_{j'n}R_{k'o}\,a_{mno} ,
\end{align}
where the indices $m,n,o$ refer to the weak field basis, and the $\ast$
denotes a sum over permutations $P$, 
$\{i^\prime,j^\prime,k^\prime\}=P\{i,j,k\}$, which gives rise to a factor of
$n!$ for $n$ identical fields.  We now proceed to obtain these couplings in an
explicit form.

At the tree level, when expressed in terms of $\lambda_i$, the derivatives 
in  Eq.~(\ref{Eq:neutral-3-coupl}) are rather simple. The trilinear  couplings
$a_{mno}$ among the weak fields $\eta_\ell$ can be written 
as~(in units of $-iv$) \cite{Choi:1999uk,Carena:2002bb}:
\begin{align}  \label{Eq:trilin-coupl-a_mno}
a_{111}&=\half(\cos\beta\, \lambda_1 + \sin\beta\, \Re\lambda_6),
\nonumber \\
a_{112}&=\half(\sin\beta\, \Re\lambda_{345} + 3\cos\beta\, \Re\lambda_6),
\nonumber \\
a_{113}&=-\half
[\cos\beta\sin\beta\, \Im\lambda_5 + (1+2\cos^2\beta)\, \Im\lambda_6],
\nonumber \\
a_{122}&=\half(\cos\beta\, \Re\lambda_{345} + 3\sin\beta\, \Re\lambda_7),
\nonumber \\
a_{123}&=-\Im\lambda_5 - \cos\beta\, \sin\beta (\Im\lambda_6 + \Im\lambda_7),
\nonumber \\
a_{133}&=\half
      \{\cos\beta (\sin^2\beta\, \lambda_1 + \cos^2\beta\, \Re\lambda_{345}
                        -2\Re\lambda_5) \nonumber \\
       &\quad +\sin\beta [(\sin^2\beta - 2 \cos^2\beta)\, \Re\lambda_6
          +\cos^2\beta\, \Re\lambda_7]\},
\nonumber \\
a_{222}&=\half(\sin\beta\, \lambda_2 + \cos\beta\, \Re\lambda_7),
\nonumber \\
a_{223}&=-\half[\cos\beta\, \sin\beta\, \Im\lambda_5
           + (\cos^2\beta + 3\sin^2\beta) \Im\lambda_7],
\nonumber \\
a_{233}&=\half
       \{\sin\beta(\cos^2\beta\, \lambda_2 + \sin^2\beta\, \Re\lambda_{345}
            -2\Re\lambda_5) \nonumber\\
       &\quad  + \cos\beta [\sin^2\beta\, \Re\lambda_6
           + (\cos^2\beta - 2\sin^2\beta)\, \Re\lambda_7]\},
\nonumber \\
a_{333}&=\half(\cos\beta\, \sin\beta\, \Im\lambda_5 
- \sin^2\beta\, \Im\lambda_6
            -\cos^2\beta\, \Im\lambda_7),
\end{align}
where
\begin{equation}
\lambda_{345} = \lambda_3 + \lambda_4 + \lambda_5. 
\end{equation}
In order to elucidate the compact notation in  Eq.~(\ref{Eq:neutral-3-coupl}),
we explicitly write the trilinear couplings  $\lambda_{111}$ and
$\lambda_{112}$ in Appendix~\ref{App:trilin_111_112}.

Rather than studying all the  trilinear couplings of the Higgs bosons,
we shall here focus on the two couplings $\lambda_{111}$ and
$\lambda_{112}$ involving the lightest Higgs  boson $H_1$. 
In the special case when $\lambda_6=\lambda_7=0$,
and if in addition $\alpha_2=\alpha_3=0$, these couplings
take the simple form
\begin{subequations}
\label{Eq:lambda:special-case}
\begin{align} \label{Eq:lambda:special-case-111}
\lambda_{111}&=-3iv\{c_1^3\, c_\beta\lambda_1+s_1^3\,s_\beta\lambda_2
\nonumber \\
&+c_1s_1(c_1\,s_\beta+s_1\, c_\beta)\Re\lambda_{345}\}, \\
\lambda_{112}&=iv\{3c_1s_1(c_1c_\beta \lambda_1-s_1 s_\beta \lambda_2)
\nonumber \\
&-[c_1^3 s_\beta-s_1^3c_\beta+2c_1s_1(c_1c_\beta-s_1s_\beta)]
\Re\lambda_{345}\}.
\label{Eq:lambda:special-case-112}
\end{align}
\end{subequations}

In this limit of $\lambda_6=\lambda_7=0$
and $\alpha_2=\alpha_3=0$, we note the following:
\begin{itemize}
\item
For small values of $\tan\beta$, both $\lambda_{111}$ 
and $\lambda_{112}$ are determined
by $\lambda_1$ and $\Re\lambda_{345}$.
This is because for small $\tan\beta$, terms
containing $s_\beta \equiv \sin\beta$ vanish. In actual practice, due to the
constraints imposed by $B$-physics, the small $\tan\beta$ limit is 
not reached. 
\item
For large values of $\tan\beta$, both $\lambda_{111}$ 
and $\lambda_{112}$ are determined by $\lambda_2$ and $\Re\lambda_{345}$.
In this case terms containing $c_\beta \equiv \cos\beta$ vanish.
\end{itemize}
Here, the combination $\Re\lambda_{345}$ is related 
to $M_{H^\pm}^2$ via
\begin{equation}
\label{Eq:mch}
M_{H^\pm}^2=\mu^2+\half v^2[\lambda_3-\Re\,\lambda_{345}
-\Re(\lambda_6+\lambda_7)].
\end{equation}

It is instructive to compare the couplings (\ref{Eq:lambda:special-case})
with the Standard Model trilinear coupling
(\ref{Eq:SM-coupling}).  While the SM trilinear coupling is given in terms 
of only one parameter, the mass $M_H$ of the Higgs boson, those of the 
2HDM depend on several parameters, and may actually pass through
zero, even for the special case studied in Eq.~(\ref{Eq:lambda:special-case}).
In the general case, we find it convenient to study the dimensionless 
ratios of the couplings
\begin{equation} \label{Eq:xi-def}
\xi_1\equiv\frac{\lambda_{111}}{\lambda_{HHH}^{\text{SM}}}, \quad \quad  \quad
\xi_2\equiv\frac{\lambda_{112}}{\lambda_{HHH}^{\text{SM}}}, 
\end{equation}
where for the reference SM coupling we use the mass $M_1$ of the lightest
Higgs boson of the 2HDM. These dimensionless ratios of the couplings
will be calculated and discussed in the following.

\subsubsection{2HDM--MSSM correspondence}

It is also useful to compare the trilinear Higgs couplings in the 2HDM
with the corresponding ones in the minimal supersymmetric standard model.  In
the MSSM,  where there is no CP violation in the Higgs sector at the
tree level, there are six trilinear Higgs couplings.  The couplings 
corresponding to $\lambda_{111}$ and $\lambda_{112}$ are 
\cite{Barger:1991ed, Djouadi:1996ah}
\begin{subequations} \label{Eq:MSSM-couplings}
\begin{align}
\lambda_{hhh}&=- \frac{3igM_Z}{2\cos\theta_W} \cos(2\alpha)\sin(\alpha+\beta),
\label{eq:lhhh}\\
\lambda_{hhH}&=- \frac{igM_Z}{2\cos\theta_W} 
[2\sin(2\alpha)\sin(\alpha+\beta) \nonumber \\
&- \cos(2\alpha)\cos(\alpha+\beta)],
\label{eq:lhhH}
\end{align}
\end{subequations}
where $M_Z$ is the mass of the $Z$, $\theta_W$ is the Weinberg angle, and
$g$ the  $SU(2)_L$ gauge coupling. The  angle $\alpha$ is the  mixing angle in
the  CP even  Higgs  sector of  the  MSSM.  In  the  limit $\alpha_2\to0$  and
$\alpha_3\to0$, it corresponds to $\alpha_1-\pi/2$ of the 2HDM.

In the MSSM the trilinear Higgs couplings are controlled by the $SU(2)_L$ gauge
coupling $g$. In contrast, the trilinear Higgs couplings in the 2HDM arise
from the quartic terms of the Higgs potential when the SM gauge symmetry is
broken. It is easy to see that in the above limit the couplings in the two
models are simply related by the following correspondence
\begin{align} \label{Eq:MSSM-correspondence}
\lambda_1=\lambda_2&\leftrightarrow\fourth(g^2+g^\prime{}^2), \nonumber \\
\Re\lambda_{345}&\leftrightarrow-\fourth(g^2+g^\prime{}^2),
\end{align}
together with
\begin{equation} \label{Eq:lambda6=lambda7=0}
\lambda_6=\lambda_7=0,
\end{equation}
where $g^\prime$ is the $U(1)$ hypercharge coupling.  We note that the
constraint (\ref{Eq:lambda6=lambda7=0}) is accessible in approach (A), but not
in (B).  Using this correspondence, together with $\alpha_1\to\alpha+\pi/2$
and
\begin{equation} \label{Eq:g-gprime}
\frac{1}{4}(g^2+g^\prime{}^2)v
=\frac{g M_Z}{2\cos\theta_W},
\end{equation}
we find
\begin{equation} \label{Eq:MSSM-2HDM}
\lambda_{111}\leftrightarrow
\lambda_{hhh},
\quad
\lambda_{112}\leftrightarrow
\lambda_{hhH}.
\end{equation}
For the relations (\ref{Eq:MSSM-correspondence}) to be satisfied, one needs
the non-trivial relations
\begin{equation} \label{Eq:lambda-conditions}
\lambda_1=\lambda_2=-\Re\lambda_{345}.
\end{equation}
When (\ref{Eq:lambda6=lambda7=0}) holds, and
\begin{equation} 
\alpha_2=\alpha_3=0, 
\end{equation}
then (more general relations can be found in Appendix~\ref{App:quartic})
\begin{align}
\lambda_1&=\frac{1}{c_\beta^2v^2}
[c_1^2M_1^2+s_1^2M_2^2-s_\beta^2\mu^2], \nonumber \\
\lambda_2&=\frac{1}{s_\beta^2v^2}
[s_1^2M_1^2+c_1^2M_2^2-c_\beta^2\mu^2], \nonumber \\
\Re\lambda_{345}&=\frac{1}{c_\beta s_\beta v^2}
c_1s_1(M_1^2-M_2^2)+\frac{\mu^2}{v^2}.
\end{align}

We note that Eq.~(\ref{Eq:lambda-conditions}) can be satisfied when
$\alpha_1\simeq\beta\simeq\pi/4$ and $M_2$ and $\mu$ are both large compared
to $M_1$:
\begin{equation} 
\frac{\sin2\alpha_1}{\sin2\beta}=\frac{M_2^2+M_1^2}{M_2^2-M_1^2}.
\end{equation}
In this limit,\footnote{With $c_\beta^2=\half+\epsilon$, one finds to lowest
order in $\epsilon$: $c_1^2=\half+\epsilon(2\mu^2/M_2^2-1)$ and
$M_1^2=2\epsilon^2(M_2^2+\mu^2-2\mu^4/M_2^2)$.}
the 2HDM will correspond to the Higgs sector of the CP-conserving MSSM.  In
fact, since the two potentials in this limit will be the same, the other
trilinear couplings in the two models will also be identical in this limit.

We note that the MSSM couplings (\ref{Eq:MSSM-couplings}), like
those of the 2HDM, vanish for certain choices of the mixing angles
$\alpha$ and $\beta$.  The exact values of parameters where they vanish
will be modified when radiative corrections are taken into account.
However, the modifications  will be only quantitative in nature. 

As a special case of the limit of no CP violation discussed above, we note
that the MSSM Higgs couplings possess an additional important property, which
is usually referred to as the decoupling~\cite{Gunion:2002zf}.  This
property can be described as follows. If $M_{A^0}\gg M_Z$, then
$\alpha\to\beta-\pi/2$ and the coupling $\lambda_{hhh}$ approaches
$\lambda_{HHH}^\text{SM}$ \cite{Hollik:2001px,Dobado:2002jz}. It follows that
in this limit, which is called decoupling limit, $\xi_1\to1$.  In the present 
notation, this requires $\alpha_1\simeq\beta$, with $M_2\simeq\mu$. 
(However, the correspondence between the 2HDM and the MSSM described 
above, fails for $\tan\beta$ away from unity.)
We will return to the issue of decoupling in Sec.~\ref{SubSec:decoupling}.
\subsubsection{Numerical study}

For fixed values of the Higgs boson masses, only a small domain in the
parameter space of $\vecalpha=(\alpha_1,\alpha_2,\alpha_3)$ is compatible with
positivity~\cite{Deshpande:1977rw,ElKaffas:2006nt}, and the
perturbative unitarity
in the Higgs sector \cite{Kanemura:1993hm,Akeroyd:2000wc,Ginzburg:2003fe}.
Furthermore, there are experimental constraints from $B$ physics: $B\to
X_s\gamma$ \cite{Koppenburg:2004fz,Barberio:2006bi, Misiak:2006zs},
$B\to\tau\bar\nu_\tau$ \cite{Ikado:2006un,Hou:1992sy}, and $B-\bar B$
oscillations \cite{Barberio:2006bi,Urban:1997gw}.  These do not depend on
details of the neutral Higgs sector, but since these processes can get
contributions from $H^\pm$ exchange, they constrain the allowed values of
$\tan\beta$ and $M_{H^\pm}$.  Furthermore, there are experimental constraints
that do depend on the neutral Higgs sector. These are $R_b$
\cite{Denner:1991ie,Yao:2006px}, non-observation of a light neutral Higgs
boson at LEP~\cite{Yao:2006px}, and most importantly the constraint
arising~\cite{:2005em,Bertolini:1985ia,ElKaffas:2006nt} from $\Delta\rho$
\cite{Grimus:2007if}.  
In contrast to the case of the MSSM, $(g-2)$ does not play any 
important role here. The experimental constraints on the Higgs sector 
of the 2HDM are discussed in \cite{ElKaffas:2007rq}. Here we shall follow the 
same approach, defining a $\chi^2$ function 
\begin{equation} \label{Eq:chi2-all}
\chi^2(\vecalpha)=\chi^2_\text{general}+\sum_i\chi_i^2(\vecalpha),
\end{equation}
where the first term, which is independent of $\vecalpha$, is due to various
$B$-physics constraints, $\bar B\to X_s\gamma$, $B^-\to\tau\bar\nu_\tau$ and
$B$--$\bar B$ oscillations
\begin{equation} \label{Eq:chi2-general}
\chi^2_\text{general}
=\chi^2_{b\to s\gamma}+\chi^2_{b\to\tau\nu}+\chi^2_{B-\bar B},
\end{equation}
and the second term is a sum over contributions due to observables ${\cal
O}_i$ that depend on $\vecalpha$.  These are the non-observation of a light
neutral Higgs boson at LEP2, the $Z^0\to b\bar b$ decay rate, and
$\Delta\rho$:
\begin{equation} \label{Eq:chi2-ind-neutr}
\chi^2_i(\vecalpha)
=\frac{({\cal O}_{i,\text{2HDM}}(\vecalpha)-{\cal O}_{i,\text{ref}})^2}
{[\sigma({\cal O}_{i})]^2}.
\end{equation}
We adopt the same definitions and experimental data as were used in
\cite{ElKaffas:2007rq}, allowing parameters for which $\chi^2\le5.99$,
corresponding to a 95\%~C.~L.

Some of the main features of these constraints are worth stressing.
First of all, for small values of $\mu^2$, the allowed range of $\tan\beta$ is 
restricted by unitarity to values below 5 -- 8 \cite{ElKaffas:2007rq}.
Secondly, for large values of $M_{H^\pm}$, the $\Delta\rho$ constraint
requires  $M_2\sim M_3\sim M_{H^\pm}$, especially for  large values
$\tan\beta$ \cite{WahabElKaffas:2007xd}.

\begin{figure}[htb]
\begin{center}
\includegraphics[width=95mm]{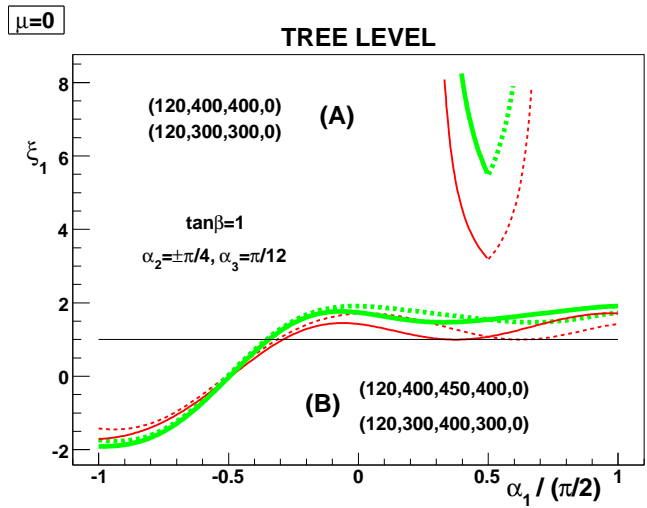}
\vspace*{-2mm}
\caption{\label{Fig:h_j11-0} The trilinear coupling ratio $\xi_1$
defined in (\ref{Eq:xi-def}) at the tree level plotted as a function of
$\alpha_1$ for both the approaches (A) and (B),  for two different values of 
$\alpha_2$ ($-\pi/4$, dashed, and $+\pi/4$, solid), both with 
$\alpha_3=\pi/12$.  The masses of the Higgs bosons
are $M_1=120~\text{GeV}$, with $M_2 = M_{H^\pm} = 300~\text{GeV}$ (red, thin
lines), and $M_2 = M_{H^\pm} = 400~\text{GeV}$ (green, heavy
lines). Furthermore, $\mu=0$ and $\tan\beta=1$.  
The numbers in insets give values of $(M_1,M_2,M_{H^\pm},\mu)$ for 
approach (A) and for those of $(M_1,M_2,M_3,M_{H^\pm},\mu)$ for approach (B).}
\end{center}
\end{figure}

Within an allowed domain in the parameter space, the trilinear couplings can
have a very strong dependence on the neutral Higgs boson mixing angles, as is
illustrated in Fig.~\ref{Fig:h_j11-0}, where we have plotted $\xi_1$ as a
function of $\alpha_1$, keeping $\alpha_2=\pm\pi/4$ and $\alpha_3=\pi/12$
fixed.

In this figure, the values of the trilinear coupling are calculated at tree
level with $M_1=120~\text{GeV}$, $\mu=0$, $\tan\beta=1$, $\alpha_2=\pm\pi/4,$
and $\alpha_3=\pi/12$.  The upper set of lines in this 
figure~(larger values of the trilinear couplings) refer to the approach (A) of
Sec.~\ref{sect:tree-level-neut-masses}, where we take $\lambda_6=\lambda_7=0$,
and fix $M_2=M_{H^\pm}=300~\text{GeV}$, or $M_2=M_{H^\pm}=400~\text{GeV}$, as
indicated in the caption.  The lower set of lines refer to the approach (B),
where we fix $(M_2,M_3,M_{H^\pm})$ equal to $(300,400,300)~\text{GeV}$, or
$(400,450,400)~\text{GeV}$, with the additional constraints $\Im\lambda_5=0$,
$\Re\lambda_6=\Re\lambda_7=0$.

From this figure we note a marked qualitative difference between the values of
the trilinear coupling calculated via the two approaches of specifying the
input parameters. In general, they both lead to solutions for only limited
ranges of $\alpha_1$, keeping $M_3$ fixed [approach~(B)], the variation of
the coupling with $\alpha_1$ is modest, whereas when $\lambda_6=\lambda_7=0$ 
and $M_3$ is calculated [approach~(A)], the variation with $\alpha_1$ is 
quite strong~(upper part of the figure).

The different behavior of the trilinear coupling in the two approaches can be
understood as follows.  The trilinear couplings are linear functions of the
quartic couplings $\lambda_i$ in the Higgs potential~(\ref{hpotential}).  In
both the approaches, (A) and (B), the couplings $\lambda_i$ are rather smooth
functions of $\alpha_1$. They are also comparable in magnitude, with the
exception that in approach (B), where we fix $M_3$, we also take as input
$\Im\lambda_5=0$. Thus, the role played by $\Im\lambda_5$ in approach (A) is
played by $\Im\lambda_6$ and $\Im\lambda_7$ in approach (B).  This switch of
role from $\Im\lambda_5$ to $\Im\lambda_6$ and $\Im\lambda_7$ has as a
consequence, for the case studied in Fig.~\ref{Fig:h_j11-0}, that
$a_{333}$~(which is quite large) switches sign. Since the coefficient in
(\ref{Eq:neutral-3-coupl}) that relates $a_{333}$ to $\lambda_{111}$ is
essentially $R_{13}^3$, with $R_{13}=\sin\alpha_2=\pm1/\sqrt{2}$, this sign
change of the large coupling $a_{333}$ is the dominant effect which causes the
difference between the parameter choices in the approaches (A) and (B) (for
the set of parameters considered). In fact, truncating
Eq.~(\ref{Eq:neutral-3-coupl}) with $a_{333}=0$, one finds the ratio of the
couplings $\xi_1=3.8-5$ for approach (A), and $3.6-3.8$ for approach (B),
where we have taken $M_1=120~\text{GeV}$, $M_2=400~\text{GeV}$,
$\alpha_2=-\pi/4$, and $\alpha_4=\pi/12$ for both the approaches.  Thus, one
may conclude that the two different ways of introducing CP violation in the
two Higgs doublet model lead to very different values for the trilinear
coupling $\lambda_{111}$.

Furthermore, we note that $\Re\lambda_5$  depends on $\alpha_1$ through 
$M_3$ only~\cite{ElKaffas:2007rq}. Thus, in approach (A), with $M_3$ 
depending on $\alpha_1$, the resulting $\Re\lambda_5$ will
also vary, whereas in approach (B), $\Re\lambda_5$ is constant. This
contributes to the stronger variation of $\xi_1$ with $\alpha_1$ in 
approach (A).

In view of the strong dependence of the trilinear couplings on the mixing
angles, we shall henceforth average over these angles within the allowed
ranges.  
In Figs.~\ref{Fig:h_j11-tree-level-cpc} and
\ref{Fig:h_j11-tree-level} we show the tree-level ratios
\begin{equation} \label{Eq:xi-def-avg}
\langle\xi_1\rangle
=\frac{\langle\lambda_{111}\rangle}{\lambda^\text{SM}_{HHH}},
\quad
\langle\xi_2\rangle
=\frac{\langle\lambda_{112}\rangle}{\lambda^\text{SM}_{HHH}},
\end{equation}
for a representative choice of parameter sets.  These are {\it averages},
obtained by scanning over the $\vecalpha=(\alpha_1,\alpha_2,\alpha_3)$ space,
subjecting all model points to positivity, unitarity, and the experimental
constraints discussed above, working within the approach (A) defined in
Sec.~\ref{sect:tree-level-neut-masses}, and normalizing them with respect to
the SM coupling, as defined in Eq.~(\ref{Eq:xi-def}).
The scans are performed over $200\times200\times100$ points.

Before considering the general case of arbitrary CP violation, we
first consider,  in Fig.~\ref{Fig:h_j11-tree-level-cpc}, a region of parameter 
space near the limit (\ref{Eq:CPcons-3}) of no CP violation:
\begin{equation} \label{Eq:no-cpv}
|\alpha_2|\leq\alpha_0, \quad
|\alpha_3|\leq\alpha_0, \quad \alpha_0=0.05\times\pi/2,
\end{equation}
referred to as {\em ``minimal CP violation''}. From the above discussion (see
Eqs.~(\ref{Eq:MSSM-correspondence})--(\ref{Eq:MSSM-2HDM})), it follows that
this corresponds to a domain of parameters close to the CP-conserving 
Higgs sector of the MSSM. We note that the approach~(A), where we determine 
$M_3$ from $M_1$, $M_2$ and the rotation matrix, does not permit us to go 
all the way to the limit $\alpha_2=\alpha_3=0$.

\begin{widetext}

\begin{figure}[bht]
\begin{center}
\includegraphics[width=150mm]{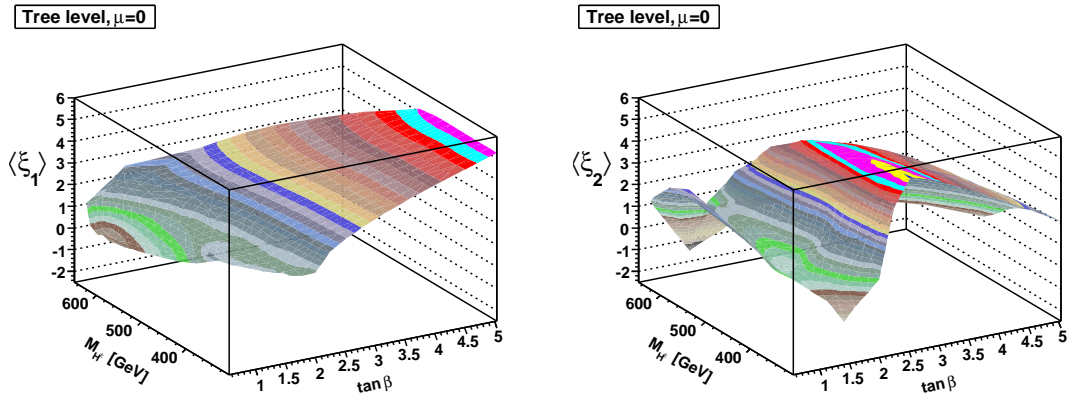}
\vspace*{-5mm}
\caption{\label{Fig:h_j11-tree-level-cpc} Trilinear Higgs coupling ratios
$\langle\xi_1\rangle$ and $\langle\xi_2\rangle$ as defined in
Eq.~(\ref{Eq:xi-def-avg}) at the tree-level, in approach (A), for the values of
the Higgs masses $M_1=120~\text{GeV}$, $M_2=300~\text{GeV}$ and $\mu=0$.  The
ratios are plotted for the case of the CP-conserving MSSM-like limit as
defined in Eq.~(\ref{Eq:no-cpv}).}
\end{center}
\end{figure}

\begin{figure}[htb]
\begin{center}
\includegraphics[width=150mm]{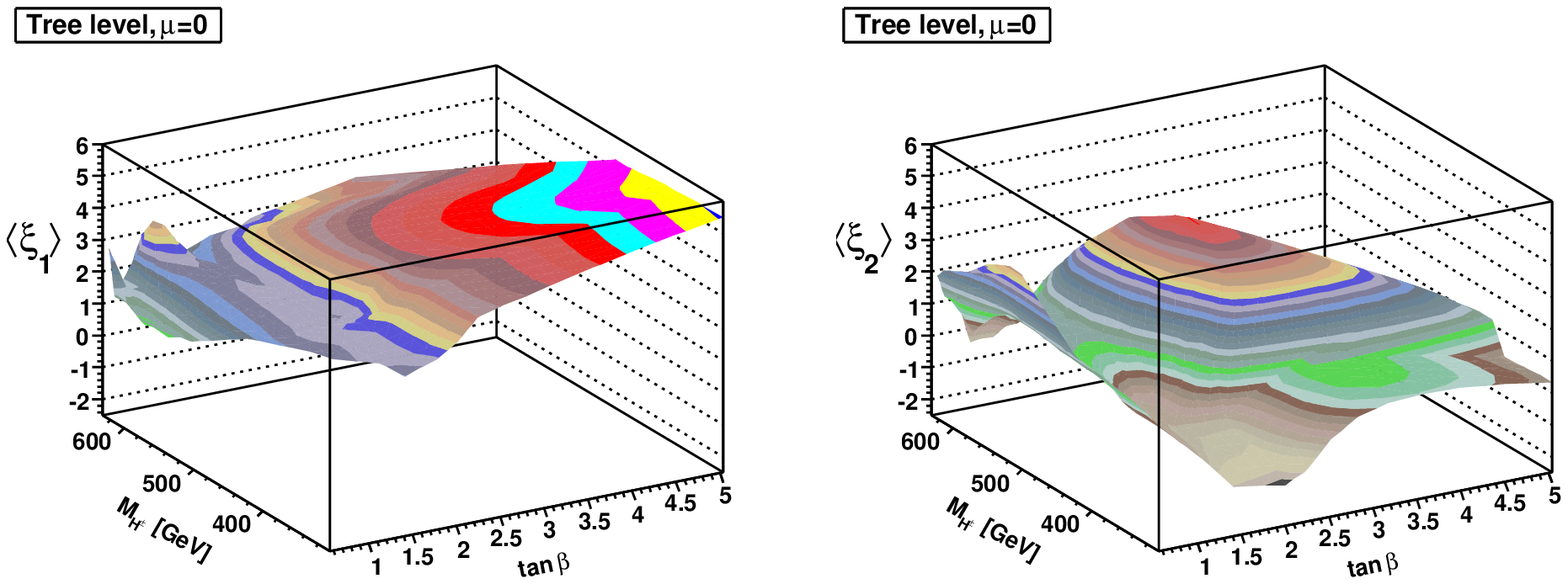}
\includegraphics[width=150mm]{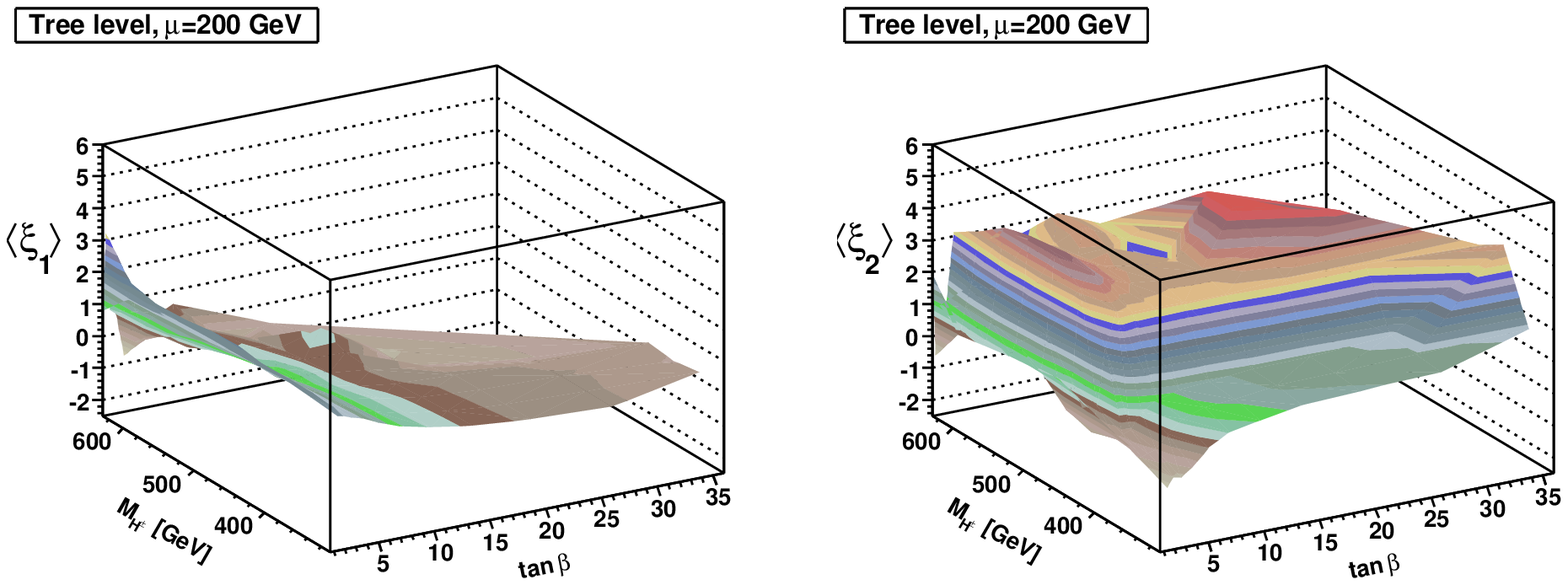}
\vspace*{-5mm}
\caption{\label{Fig:h_j11-tree-level} Tree-level trilinear coupling ratios
$\langle\xi_1\rangle$ and $\langle\xi_2\rangle$ as defined in
Eq.~(\ref{Eq:xi-def-avg}), in approach (A),  for the values of the Higgs masses
$M_1=120~\text{GeV}$ and $M_2=300~\text{GeV}$.  For the top panel $\mu=0$,
whereas for the bottom panel $\mu=200~\text{GeV}$. This plot is for the
general case including CP violation, and is compatible with general
constraints on the model.}
\end{center}
\end{figure}

In this figure, which is valid for $\mu=0$, the plots do not extend much
beyond $\tan\beta=5$.  This is caused by the constraint of unitarity in the
Higgs--Higgs scattering sector \cite{WahabElKaffas:2007xd}.  At the higher
values of $\tan\beta$, there is also a narrowing of the allowed region,
largely due to the constraints following from $\Delta\rho$ and
the $B$ physics~\cite{ElKaffas:2006nt,WahabElKaffas:2007xd}.

The surfaces representing $\langle\xi_1\rangle$ and $\langle\xi_2\rangle$ in
Fig.~\ref{Fig:h_j11-tree-level-cpc} are in fact quite remininscent of the
corresponding averages of $\lambda_1$ and $\Re\lambda_{345}$, respectively.
The details of these correspondences depend on which domains of the
$\vecalpha$ space (in this case, $\alpha_1$) is populated. Typically,
only scattered regions are allowed \cite{ElKaffas:2006nt}.

On the other hand, in Fig.~\ref{Fig:h_j11-tree-level} we have scanned 
over the full range of values
of $\alpha_2$ and $\alpha_3$, compatible with all the constraints.

From these figures, we see that in much of the $\tan\beta-M_{H^\pm}$ parameter
space, there is a considerable enhancement of the trilinear couplings as
compared with the corresponding trilinear coupling in the SM. This is
consistent with the special case studied in Fig.~\ref{Fig:h_j11-0}.  

Incorporating CP violation in the 2HDM, we see a considerable change in the
behavior of the couplings as we go from Fig.~\ref{Fig:h_j11-tree-level-cpc} to
Fig.~\ref{Fig:h_j11-tree-level}. This reflects the fact seen in
Fig.~\ref{Fig:h_j11-0} that the couplings have a strong dependence on the
mixing angles, and, thus, on how CP violation is incorporated in the model.

Furthermore, there is a rapid variation of the Higgs coupling ratios
around $\tan\beta={\cal O}(2)$.  As we shall see in the following, this is 
accompanied by a strong variation as one scans across the different 
values of $\vecalpha$.


\begin{figure}[htb]
\begin{center}
\includegraphics[width=150mm]{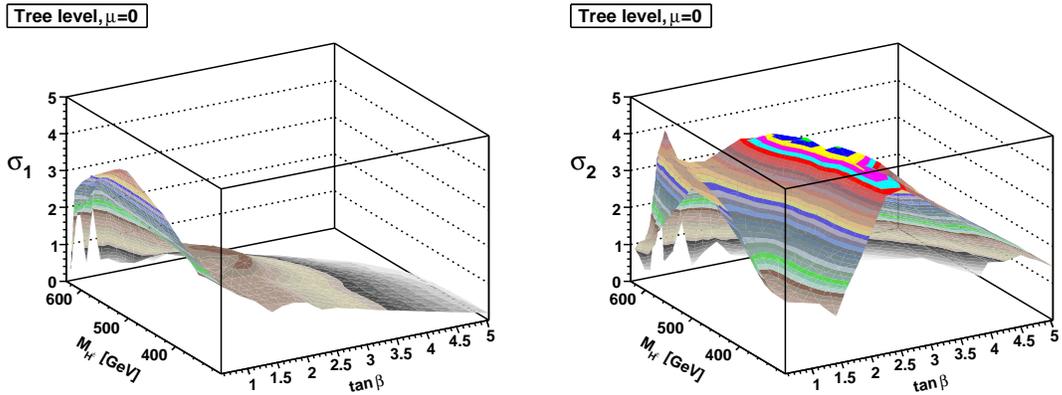}
\vspace*{-5mm}
\caption{\label{Fig:var-tree-level} Variance of  trilinear
coupling ratios $\xi_1$ and $\xi_2$ at the tree level. The parameters here 
correspond to the upper panels ($\mu=0$) of Fig.~\ref{Fig:h_j11-tree-level}.}
\end{center}
\end{figure}

\begin{figure}[htb]
\begin{center}
\includegraphics[width=150mm]{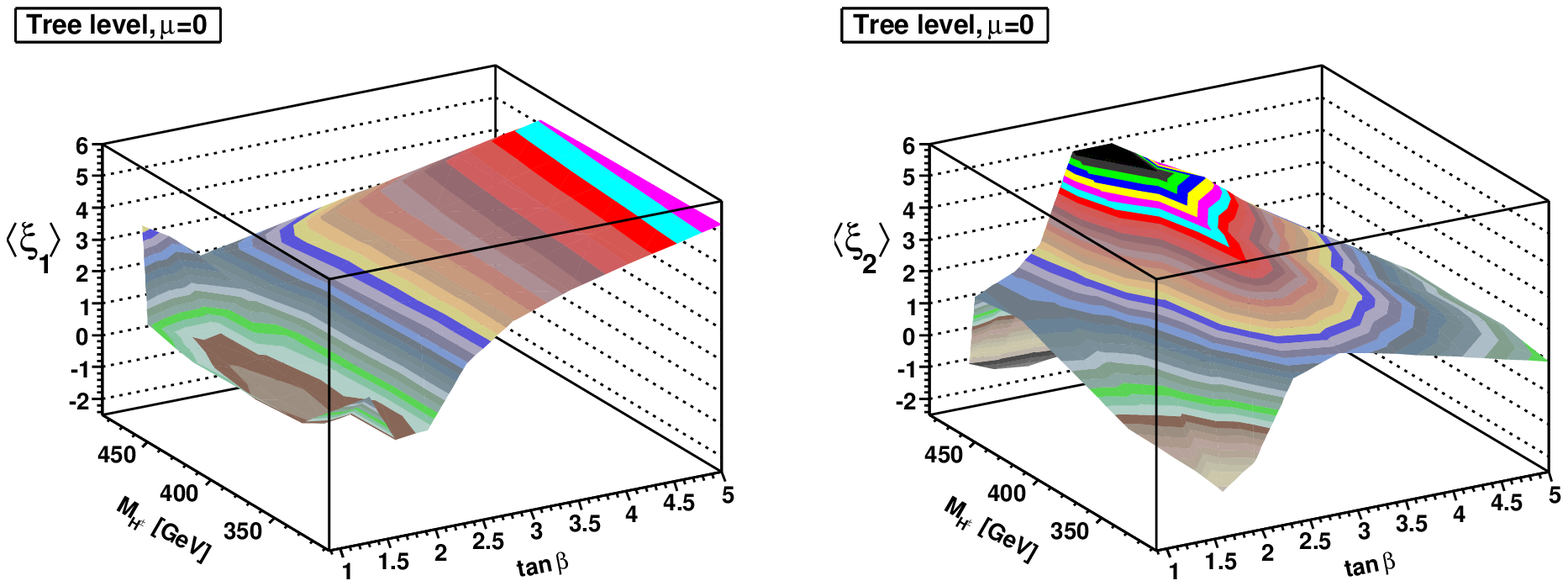}
\vspace*{-5mm}
\caption{\label{Fig:F.h_j11-tree-level} Trilinear coupling ratios
$\langle\xi_1\rangle$ and $\langle\xi_2\rangle$ at the tree level~[see
Eq.~(\ref{Eq:xi-def-avg})], for approach (B),  with $M_1=120~\text{GeV}$,
$M_2=300~\text{GeV}$, $M_3=400~\text{GeV}$ and $\mu=0$.}
\end{center}
\end{figure}

The trilinear coupling has a rather complicated behaviour across the
$\tan\beta$--$M_{H^\pm}$ plane, despite the smoothening that is implicit in
the averaging over $\vecalpha$.  As is evident from Fig.~\ref{Fig:h_j11-0} for
$\xi_1$, the ratios $\xi_1$ and $\xi_2$ of trilinear couplings also vary
considerably over the parameter space of $\vecalpha$.  One measure of this
variation is the {\it variance}, defined as
\begin{equation}
\sigma_i=\sqrt{\langle(\xi_i-\langle\xi_i\rangle)^2\rangle}.
\end{equation}

In Fig.~\ref{Fig:var-tree-level} we display the variances $\sigma_1$ and
$\sigma_2$ corresponding to the upper panels of
Fig.~\ref{Fig:h_j11-tree-level}.  This quantity (in units of the SM trilinear
coupling) is seen to be quite considerable, in particular for $\xi_2$ and
for moderate values of $\tan\beta$.

For approach (B), when we keep all three neutral Higgs boson masses fixed, the
behaviour is smoother, as seen in Fig.~\ref{Fig:F.h_j11-tree-level}.  For
the parameters considered (note that $\mu=0$), the values of $\tan\beta$
extend up to about 5.  We note that the quantity $\langle\xi_1\rangle$ is
rather similar to that shown in Figs.~\ref{Fig:h_j11-tree-level-cpc} and
\ref{Fig:h_j11-tree-level}, increasing more or less monotonically with
$\tan\beta$.  The ratio $\langle\xi_2\rangle$ differs more, exhibiting a bulge
of rather strong $H_1H_1H_2$ coupling for $\tan\beta=2-3$ and
$M_{H^\pm}=450-500~\text{GeV}$.

\subsection{Charged--Charged--Neutral Higgs Couplings}
Although in this paper we are mainly concerned with the trilinear couplings 
of the neutral Higgs bosons, here we briefly discuss  the trilinear couplings 
of the charged Higgs boson of the 2HDM~\cite{Carena:2002bb}. These are
given by
\begin{equation}  \label{Eq:charged-coupl}
\lambda_{i+-}=\sum_{m=1,2,3}R_{im}\,b_m.
\end{equation}
We note that the index $m$ (like the indices $n$ and $o$ in 
Eq.~(\ref{Eq:neutral-3-coupl})) refers to the weak interaction 
eigenstates~($\eta_1$, $\eta_2$ and $\eta_3$) of Eq.~(\ref{Eq:etas}).  
The coefficients $b_m$  in (\ref{Eq:charged-coupl}) 
can be written as~(in units of $-iv$):
\begin{align}
b_1&=\cos\beta\{\sin^2\beta\, (\lambda_1 - \lambda_4 - \Re\lambda_5)
+ \cos^2\beta\, \lambda_3 \nonumber \\
&+ \cos\beta\, \sin\beta\, [(\tan^2\beta - 2)\, \Re\lambda_6 + \Re\lambda_7]\},
\nonumber \\
b_2&=\sin\beta\{\cos^2\beta\, (\lambda_2 - \lambda_4 - \Re\lambda_5)
+ \sin^2\beta\, \lambda_3 \nonumber \\
&+\cos\beta\, \sin\beta\, [\Re\lambda_6 + (\cot^2\beta -2)\, \Re\lambda_7]\},
\nonumber \\
b_3&=\cos\beta\, \sin\beta\,\Im\lambda_5 - \sin^2\beta\,\Im\lambda_6
- \cos^2\beta\, \Im\lambda_7.
\end{align}

In Fig.~\ref{Fig:h_j11-0-pm} we display the coupling of $H_1$ to the charged
$H^+H^-$ pair, for the same choice of parameters as in Fig.~\ref{Fig:h_j11-0}.
We note that in approach (B), with $\tan\beta=1$ and $\Im\lambda_5=0$, this
coupling has two interesting features: (1) It does not depend on the sign of
$\alpha_2$; and (2) it passes through zero for $\alpha_1=-\pi/4$.  The
independence from the sign of $\alpha_2$ can be understood in the
following manner. For $i=1$, the first two terms in 
Eq.~(\ref{Eq:charged-coupl}) (proportional to
$R_{11}$ and $R_{12}$) contribute terms which are odd in $\alpha_2$, arising
from $\lambda_1$, $\lambda_2$ and $\lambda_3$. These odd parts are
proportional to $(M_3^2-M_2^2)$, and cancel against odd parts coming from the
third term in (\ref{Eq:charged-coupl}), which are proportional to
$\Im\lambda_6+\Im\lambda_7$, which in turn is proportional to ${\cal
M}_{13}+{\cal M}_{23}$. 

Furthermore, the vanishing of the coupling at
$\alpha_1=-\pi/4$ can be understood as follows. In approach (B), with
$\tan\beta=1$ and $\alpha_1=-\pi/4$, we have (see Eqs.~(3.1) and (3.2) of
Ref.~\cite{ElKaffas:2007rq})
\begin{align}
\lambda_1(\alpha_2,\alpha_3)
&=\frac{1}{v^2}[c_2^2M_1^2
+(c_3-s_2s_3)^2M_2^2 \nonumber \\
&+(s_3+s_2c_3)^2M_3^2-\mu^2], \nonumber \\
\lambda_2(\alpha_2,\alpha_3)
&=\frac{1}{v^2}[c_2^2M_1^2
+(c_3+s_2s_3)^2M_2^2 \nonumber \\
&+(s_3-s_2c_3)^2M_3^2-\mu^2].
\end{align}
Thus, for these particular values of  $\tan\beta$ and $\alpha_1$,
$\lambda_2(-\alpha_2,\alpha_3) =\lambda_1(\alpha_2,\alpha_3)$, the first two
terms in Eq.~(\ref{Eq:charged-coupl}) 
cancel, and we are left with
\begin{equation}
\lambda_{1+-}=R_{13}b_3
=\half s_2(\Im\lambda_5-\Im\lambda_6-\Im\lambda_7)=0,
\end{equation}
where in the last step we have used  $\Im\lambda_5=0$,
and the relations (see Eq.~(\ref{Eq:M^2_ij}))
\begin{equation}
\Im\lambda_5+\Im\lambda_6+\Im\lambda_7
=-\frac{\sqrt{2}}{v^2}[{\cal M}_{13}^2+{\cal M}_{23}^2],
\end{equation}
valid for $\tan\beta=1$, and
\begin{equation}
{\cal M}_{13}^2+{\cal M}_{23}^2=0,
\end{equation}
valid for $\alpha_1=-\pi/4$.

\begin{figure}[htb]
\begin{center}
\includegraphics[width=95mm]{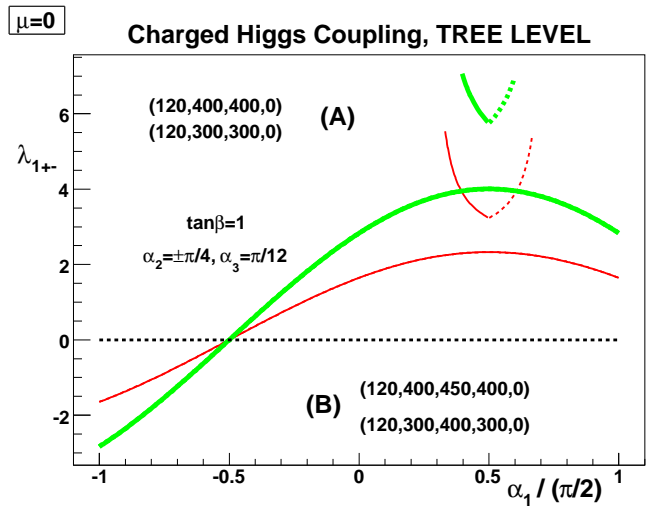}
\vspace*{-2mm}
\caption{\label{Fig:h_j11-0-pm} The trilinear coupling
$\lambda_{1+-}$ at the tree level as a function of $\alpha_1$, 
for two different values of
$\alpha_2$ ($-\pi/4$, dashed, and $+\pi/4$, solid), both with
$\alpha_3=\pi/12$.  The masses and parameters are the same as in
Fig.~\ref{Fig:h_j11-0}.}
\end{center}
\end{figure}

In Fig.~\ref{Fig:lambda_kpm-tb-mh_ch}, we show the coupling of the lightest
neutral Higgs bosons to the charged Higgs bosons, averaged over $\vecalpha$,
in approach~(A).  Corresponding to what we observed in
Fig.~\ref{Fig:h_j11-0-pm}, these averages over $\vecalpha$ also vanish along
certain lines in the parameter space.


\begin{figure}[htb]
\begin{center}
\includegraphics[width=150mm]{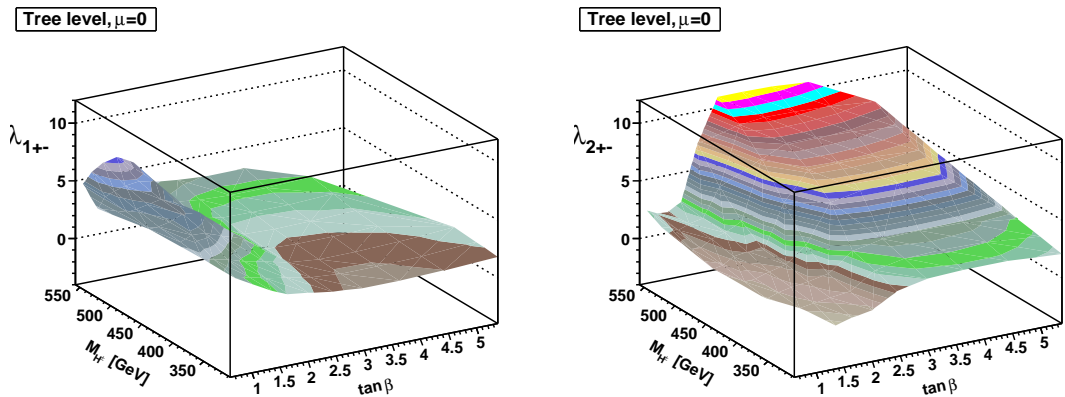}
\vspace*{-5mm}
\caption{\label{Fig:lambda_kpm-tb-mh_ch} Tree-level trilinear couplings
$\lambda_{1+-}$ and $\lambda_{2+-}$ (in units of $-iv$) of the two lightest
neutral Higgs bosons, $H_1$ and $H_2$, to a charged pair for
$M_1=120~\text{GeV}$, $M_2=300~\text{GeV}$ and $\mu=0$.
The couplings are averaged over $\vecalpha$, and have been plotted for 
approach (A).}
\end{center}
\end{figure}
\end{widetext}

\section{One-loop Corrections to the Higgs Boson Masses}
\label{sect:one-loop-level-masses}

\subsection{One-loop effective  potential}

For a realistic computation of the Higgs boson properties in the 2HDM, we
must take into account the one-loop radiative corrections to the tree-level
potential~(\ref{hpotential}).  When this is done, the relations among masses
and the couplings of the tree-level potential undergo a change.  In the MSSM,
this effect, which is dominated by the top~(and stop) quark contributions, is
known to be very important~\cite{Okada:1990vk} (see also
Ref.~\cite{Ham:2002ps}). In the 2HDM, there being no one-loop contribution
from squarks, the main contributions at one loop are generated by the Higgs
bosons.  We shall here study the one-loop radiative corrections to the neutral
Higgs sector using the method of one-loop effective
potential~\cite{Coleman:1973jx}. The one-loop corrections to the tree-level
potential~(\ref{hpotential}) are given by
\begin{align} \label{Eq:DeltaV}
\Delta V
&=\frac{1}{64\pi^2}\biggl[\sum_{\text{bosons}}
M^4\biggl(\log\frac{M^2}{Q^2}-\frac{3}{2}\biggr)
\nonumber \\
&-\sum_{\text{fermions}}
M^4\biggl(\log\frac{M^2}{Q^2}-\frac{3}{2}\biggr)\biggr],
\end{align}
where the sums run over all bosonic and fermionic degrees of freedom,
respectively, and $Q^2$ represents the scale at which the couplings are
evaluated.  As indicated above, the dominant contributions 
to (\ref{Eq:DeltaV}) come from the Higgs bosons~(neutral and charged), 
and the top quarks, all of which we will  calculate in this section. 
Taking into account the spin, charge and color degrees of freedom, we obtain 
\begin{align} \label{Eq:V-loop}
\Delta V&=\Delta V_\text{neutral Higgs}
+\Delta V_\text{charged Higgs}+\Delta V_\text{top} \nonumber \\
&=\frac{1}{64\pi^2}\sum_{\ell=1}^3
\biggl[M_\ell^4\biggl(\log\frac{M_\ell^2}{Q^2}-\frac{3}{2}\biggr)\biggr]
\nonumber \\
&+\frac{1}{32\pi^2}
M_{H^\pm}^4\biggl(\log\frac{M_{H^\pm}^2}{Q^2}-\frac{3}{2}\biggr)
\nonumber \\
&-\frac{3}{16\pi^2}\, m_t^4\biggl(\log\frac{m_t^2}{Q^2}-\frac{3}{2}\biggr),
\end{align}
where all masses are understood to be field-dependent masses.  Thus, after 
differentiation analogous to the one in Eq.~(\ref{Eq:M_sq-def}) for the
tree-level potential, for example, we do {\it not} set
the fields $\eta_1$, $\eta_2$ and $\eta_3$ to zero, but instead evaluate
\begin{equation} 
{\cal M}^2_{ij}(\eta_1,\eta_2,\eta_3)
=\frac{\partial}{\partial \eta_i}
\frac{\partial}{\partial \eta_j}\,V(\eta_1,\eta_2,\eta_3,H^\pm,G^\pm),
\end{equation}
and similarly
\begin{equation}
M_{H^\pm}^2(\eta_1,\eta_2,\eta_3) 
=\frac{\partial}{\partial H^+}
\frac{\partial}{\partial H^-}\,V(\eta_1,\eta_2,\eta_3,H^\pm,G^\pm),
\end{equation}
and after having performed the differentiation, we set
$H^{\pm} = G^{\pm} = 0.$

At tree level, the three minimization conditions (\ref{Eq:min-cond-tree}) (two
real, and one imaginary) can be used to eliminate the bilinear parameters
$m_1^2$ and $m_2^2$, as well as relate $\Im\lambda_5$ to $\Im m_{12}^2$. At
the one-loop level, the expressions for $m_1^2$ and $m_2^2$ are modified due
to the contributions coming from $\partial\Delta V/\partial\eta_1$ and
$\partial\Delta V/\partial\eta_2$.  This amounts to adjusting the parameters
of the potential such that the minimum is stationary in terms of the same
vacuum expectation values $v_1$ and $v_2$ as for the tree-level potential
(i.e., $\tan\beta$ remains unchanged). In the following we shall calculate the
different contributions to the neutral Higgs boson masses arising from
(\ref{Eq:V-loop}) using the above procedure.

\subsection{One-loop corrections to Higgs  masses}

The mass squared matrix for the neutral Higgs bosons will receive
contributions from one-loop effects due to neutral Higgs bosons, which can be
written as
\begin{align} 
\frac{\partial^2 \Delta V_\text{neutral Higgs}}{\partial \eta_i\partial \eta_j}
&=\frac{1}{32\pi^2}\sum_{\ell=1}^3
\biggl[\log\frac{M_\ell^2}{Q^2}
\frac{\partial M_\ell^2}{\partial \eta_i}
\frac{\partial M_\ell^2}{\partial \eta_j}
\nonumber \\
&+M_\ell^2\biggl(\log\frac{M_\ell^2}{Q^2}-1\biggr)
\frac{\partial^2 M_\ell^2}{\partial \eta_i \partial \eta_j}\biggr].
\end{align}
The derivatives of $M_\ell^2$ with respect to the fields $\eta_i$,
which are rather involved, are
discussed in detail in Appendix~\ref{App:cubic}.

The corresponding second derivatives of the charged-Higgs-field-dependent part
of the one-loop corrections to the potential,  $\Delta V_\text{charged Higgs}$,
in (\ref{Eq:V-loop}) are given by
\begin{align}
\frac{\partial^2 \Delta V_\text{charged Higgs}}
{\partial \eta_i\partial \eta_j} 
&=\frac{1}{16\pi^2}
\biggl[\log\frac{M_{H^\pm}^2}{Q^2} \frac{\partial M_{H^\pm}^2}{\partial
\eta_i} \frac{\partial M_{H^\pm}^2}{\partial \eta_j}
\nonumber \\
&+M_{H^\pm}^2\biggl(\log\frac{M_{H^\pm}^2}{Q^2}-1\biggr) \frac{\partial^2
M_{H^\pm}^2}{\partial \eta_i \partial \eta_j}\biggr].
\end{align}

Finally, there are one-loop corrections due to the top quark, which can be
written as
\begin{equation}
\frac{\partial^2 \Delta V_\text{top}}{\partial \eta_i\partial \eta_j}
=-\frac{3}{2\pi^2}\,\frac{m_t^4}{v_2^2}\biggl(2\log\frac{m_t^2}{Q^2}-1\biggr)
\delta_{i2}\delta_{j2},
\end{equation}
where in the last equation we have used $\partial
m_t/\partial\eta_i=(m_t/v_2)\delta_{i2}$. We neglect the contributions from
lighter fermions.

The full one-loop  mass-squared matrix for the neutral Higgs bosons can, thus, 
be written as a sum of four terms:
\begin{align} \label{Eq:mass-sq-1loop}
{\cal M}^2_{ij}(\eta_1,\eta_2,\eta_3)
&={\cal M}^2_{\text{tree},ij}(\eta_1,\eta_2,\eta_3)
+\frac{\partial^2 \Delta V_\text{neutral Higgs}}
{\partial \eta_i\partial \eta_j}
\nonumber \\
&+\frac{\partial^2 \Delta V_\text{charged Higgs}}
{\partial \eta_i\partial \eta_j} 
+\frac{\partial^2 \Delta V_\text{top}}{\partial \eta_i\partial \eta_j}.
\end{align}
From the mass-squared matrix (\ref{Eq:mass-sq-1loop}),
the eigenvalues $M_\ell(\eta_1,\eta_2,\eta_3)$ can be determined. These
are the one-loop corrected masses for the neutral Higgs bosons.
\begin{widetext}

In Fig.~\ref{Fig:mmm-orders} we show, for approach~(A), the one-loop corrected
Higgs boson masses, corresponding to the trilinear couplings studied in
Fig.~\ref{Fig:h_j11-0} (i.e., with $\alpha_2$ and $\alpha_3$ held fixed), due
to the one-loop radiative corrections discussed above.  The superscript {\rm
`LO'} here denotes lowest-order values.  The Higgs boson masses are determined
as eigenvalues of the matrix (\ref{Eq:mass-sq-1loop}), evaluated for
$\eta_1=\eta_2=\eta_3=0$.  The full $\eta$-dependence, to be discussed in the
next section, will however be required for evaluating the one-loop trilinear
Higgs couplings.

\begin{figure}[htb]
\begin{center}
\includegraphics[width=95mm]{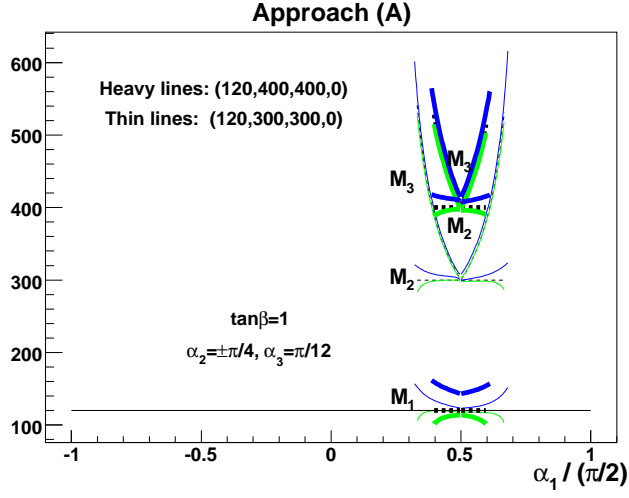}
\vspace*{-2mm}
\caption{\label{Fig:mmm-orders} One-loop Higgs masses $M_1$, $M_2$ and $M_3$
(in units of GeV) plotted as functions of $\alpha_1$, with
$\alpha_2=\pm\pi/4$, both with $\alpha_3=\pi/12$.  The lowest-order Higgs
masses are $M_1^\text{LO}=120~\text{GeV}$, with
$M_2^\text{LO}=M_{H^\pm}^\text{LO}=300~\text{GeV}$ (thin lines), and
$M_2^\text{LO}=M_{H^\pm}^\text{LO}=400~\text{GeV}$ (heavy lines). Here $\mu=0$
and $\tan\beta=1$.  Dashed lines refer to tree-level values, whereas green and
blue lines refer to one-loop-corrected results, with the scale parameter $Q =
2M_1^\text{LO}$ and $Q = 500~\text{GeV}$, respectively.}
\end{center}
\end{figure}

The following points are worth noting here: (i) The tree-level value of
$M_3$, namely $M_3^\text{LO}$, has a strong dependence on $\alpha_1$.  
This is a consequence of keeping $M_1^\text{LO}$, $M_2^\text{LO}$, 
$\alpha_2$ and $\alpha_3$ fixed.  (ii) The one-loop corrections have a 
significant dependence on the scale at which they are evaluated, as is 
illustrated by comparing the values of the Higgs boson masses at two 
different values of $Q^2$.  (iii) The one-loop
correction to the Higgs boson mass can be either positive or negative.  
(iv) At some of the boundaries of the allowed parameter space, 
these one-loop corrections become large.

The reason for the strong variation of $M_3$ at the edges of the allowed
parameter space can be found from an inspection of the different origin of
such edges. These can be described as follows:
\begin{enumerate}
\item
In approach (A), at tree level, $(M_3^\text{LO})^2$ is determined as a
rational fraction, where the numerator is linear in $(M_1^\text{LO})^2$ and
$(M_2^\text{LO})^2$ (multiplying rotation matrix elements and $\tan\beta$),
whereas the denominator is a linear function in rotation matrix elements and
$\tan\beta$~(For an explicit formula, see Eq.~(4.16) of
Ref.~\cite{Khater:2003wq}.).  For critical values of the rotation matrix
elements, e.g., when $R_{31}=R_{32}\tan\beta$,
the denominator will pass through zero. 
At such points,
$M_3^\text{LO}$ diverges, and the allowed parameter space terminates.
\item
In approach (A), $M_3^2$ may at some critical values of the rotation matrix
elements become equal to $M_2^2$.  This happens when either $R_{13}=0$ or
$R_{12}\tan\beta=R_{11}$ \cite{Khater:2003wq}.  In the former case we have
$\alpha_2=0$ ($\alpha_1$ and $\alpha_3$ being arbitrary), whereas in the
latter we have $\tan\beta=\cot\alpha_1$ ($\alpha_2$ and $\alpha_3$ being
arbitrary).
\item
Positivity may break down.
\item
Perturbative unitarity may break down.
\item
Parameter space may be truncated by some experimental constraint.
\end{enumerate}
The first two of these cases are illustrated in Fig.~\ref{Fig:mmm-orders}.
We actually exclude, as `unphysical', regions where $M_1<100~\text{GeV}$ (for
$M_1^\text{LO}=120~\text{GeV}$) and where either
$M_3^\text{LO}>2M_2^\text{LO}$ or $M_3>2M_2$.


\begin{figure}[htb]
\begin{center}
\includegraphics[width=160mm]{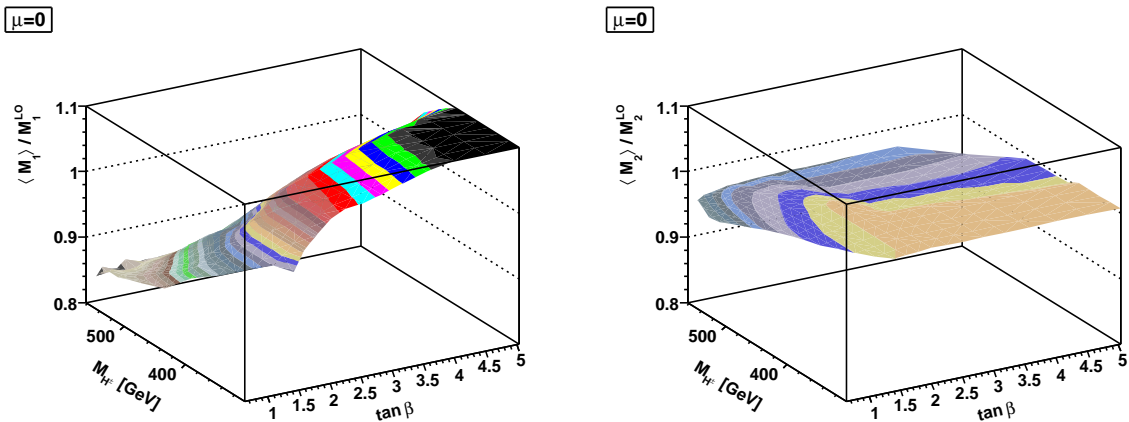}
\vspace*{-2mm}
\caption{\label{Fig:mass-corr} One-loop multiplicative mass correction for
tree-level mass parameters $M_1=120~\text{GeV}$, $M_2=300~\text{GeV}$
and $\mu=0$.}
\end{center}
\end{figure}

For approach (A), positivity is expressed in terms of simple inequalities,
whereas for approach (B), with $\alpha_6$ and $\alpha_7$ non-zero,
an involved numerical test is required \cite{ElKaffas:2006nt}.
Perturbative unitarity in the Higgs--Higgs scattering sector
tends to truncate high values of $\tan\beta$, unless $\mu$ is large
(comparable to $M_2$) \cite{ElKaffas:2007rq}.

Having described the radiative corrections to the Higgs masses, 
we next show in Fig.~\ref{Fig:mass-corr} typical one-loop multiplicative mass
corrections, i.e., $M_1/M_1^\text{LO}$ and $M_2/M_2^\text{LO}$ for the case
$M_1^\text{LO}=120~\text{GeV}$, $M_2^\text{LO}=300~\text{GeV}$, and $\mu=0$.
(This will be discussed in more detail in Sec.~\ref{sect:Summary}.)
The renormalization scale has been chosen as $Q^2=(2M_1^\text{LO})^2$.  As in
Section \ref{sect:tree-level-couplings}, an average over allowed points in
$\vecalpha$ has been performed.

For $M_1$, the loop correction is (for $Q^2=(2M_1^\text{LO})^2$) negative at
low values of $\tan\beta$, whereas it is positive for higher values.  For
$M_2$, on the other hand, the correction is smaller, and has no strong
dependence on $\tan\beta$.  The negative correction for $M_1$ is different from
the MSSM, where the one-loop corrections, dominated by the top-quark (squark)
contribution, is positive. Here, it is the Higgs sector which dominates, and
this contribution has the opposite sign, because of the bosonic nature of the
contributing particles in the loop. In actual practice the situation is
more complicated, since different Higgs fields may contribute with 
different signs, depending on how the mass of the respective Higgs 
boson compares with the scale $Q$.

In view of the fact that the one-loop correction to $M_1$ can
be significant, and negative, we impose the constraint of LEP2
non-discovery of a Higgs boson on the one-loop corrected 
lightest Higgs boson mass.  
Higgs boson masses below the ``magic'' 114.4~GeV value are of course allowed,
provided the coupling to the $Z$ and $b\bar b$ (or $\tau$ pairs)
are suppressed.

Treating the loop corrections as a perturbation, we have here imposed
the theoretical and experimental constraints on the tree-level
parameters. We note that at low values of $\tan\beta$, the mass
of the lightest Higgs boson, $M_1$, may be considerably reduced.
When we proceed to study the trilinear couplings, we shall therefore
impose the LEP2 non-discovery constraint on the loop-corrected
mass, $M_1$. 
For a higher scale parameter, though, the mass may increase also
at low values of $\tan\beta$.

\section{One-loop Corrections to the  Trilinear Higgs Couplings}
\label{sect:one-loop-level-couplings}
The one-loop corrected trilinear Higgs couplings are given by the
third order derivatives of the one-loop effective potential with respect 
to different fields.  We shall separately consider the contributions to 
the potential from the neutral Higgs fields,
the charged Higgs fields and the top quark fields. We  start the
calculation by evaluating  the derivatives with respect to the weak fields, 
which can then be converted to derivatives with respect to the fields
corresponding to the mass eigenstates with the help of Eq.~(\ref{Eq:h-R-eta}).

The starting point of the calculation of the one-loop corrections to the
trilinear Higgs couplings is the expression (\ref{Eq:V-loop}). From this
we obtain  the one-loop radiative corrections to the trilinear
Higgs couplings due to  neutral Higgs bosons as 
\begin{align} \label{Eq:delta-pot-neut}
&\frac{\partial^3 \Delta V_\text{neutral Higgs}}
{\partial \eta_i\partial \eta_j\partial \eta_k}
=\frac{1}{32\pi^2}\sum_{\ell=1}^3
\biggl[\frac{1}{M_\ell^2}
\frac{\partial M_\ell^2}{\partial \eta_i}
\frac{\partial M_\ell^2}{\partial \eta_j}
\frac{\partial M_\ell^2}{\partial \eta_k} \nonumber \\
&\quad
+\log\frac{M_\ell^2}{Q^2}
\biggl(
 \frac{\partial^2 M_\ell^2}{\partial \eta_i\partial \eta_j}
 \frac{\partial M_\ell^2}{\partial \eta_k}
+\frac{\partial^2 M_\ell^2}{\partial \eta_j\partial \eta_k}
 \frac{\partial M_\ell^2}{\partial \eta_i}
\nonumber \\
&\quad
+\frac{\partial^2 M_\ell^2}{\partial \eta_k\partial \eta_i}
 \frac{\partial M_\ell^2}{\partial \eta_j}
\biggr) 
+M_\ell^2\biggl(\log\frac{M_\ell^2}{Q^2}-1\biggr)
\frac{\partial^3 M_\ell^2}
{\partial \eta_i \partial \eta_j \partial \eta_k}\biggr].
\end{align}
The derivatives of squared masses with respect to the weak fields $\eta_i$
in (\ref{Eq:delta-pot-neut})  are evaluated and discussed
in Appendix~\ref{App:cubic}.

To find one-loop charged Higgs boson correction to the trilinear couplings,
we start with  {\it field-dependent} squared masses, 
$M_{H^\pm}^2(\eta_1,\eta_2,\eta_3)$, for the charged Higgs boson. 
Noting that the potential is a function of masses, i.e.,
$\Delta V_\text{charged Higgs}=\Delta V(\eta_1,\eta_2,\eta_3)$, 
we obtain

\begin{align}
&\frac{\partial^3 \Delta V_\text{charged Higgs}} {\partial
\eta_i\partial \eta_j\partial \eta_k} 
=\frac{1}{16\pi^2}
\biggl[\frac{1}{M_{H^\pm}^2} \frac{\partial M_{H^\pm}^2}{\partial
\eta_i} \frac{\partial M_{H^\pm}^2}{\partial \eta_j}
\frac{\partial M_{H^\pm}^2}{\partial \eta_k} \nonumber \\
&\quad +\log\frac{M_{H^\pm}^2}{Q^2} \biggl(
 \frac{\partial^2 M_{H^\pm}^2}{\partial \eta_i\partial \eta_j}
 \frac{\partial M_{H^\pm}^2}{\partial \eta_k}
+\frac{\partial^2 M_{H^\pm}^2}{\partial \eta_j\partial \eta_k}
 \frac{\partial M_{H^\pm}^2}{\partial \eta_i}
\nonumber \\
&\quad +\frac{\partial^2 M_{H^\pm}^2}{\partial \eta_k\partial \eta_i}
 \frac{\partial M_{H^\pm}^2}{\partial \eta_j}
\biggr)\biggr].
\end{align}

For the one-loop corrections arising from the top-quark, we note that in
Model~II (which we consider) the up-type quarks couple only to $\Phi_2$ (and
not to $\Phi_1$). Thus, the one-loop contribution due to top quark affects
only the fields $\eta_2$.  We can then write
\begin{equation} \label{Eq:delta-pot-top}
\frac{\partial^3 \Delta V_\text{top}} 
{\partial \eta_i\partial \eta_j\partial \eta_k}
=-\frac{12}{\pi^2}\frac{m_t^4}{v_2^{3}}\log\frac{m_t^2}{Q^2}\,
\delta_{i2}\delta_{j2}\delta_{k2}.
\end{equation}

\section{Quantitative results for trilinear Higgs couplings}
\label{sec:quantitative results}

We have seen that the lowest order trilinear couplings are highly dependent on
the details of the two Higgs doublet model. This sensitivity on the details of
the model is actually enhanced when we take into account the one-loop
corrections to the Higgs potential.  The loop corrections to the trilinear
couplings can vary strongly with the parameters of the model, and also have a
considerable dependence on the scale $Q^2$ used in
Eqs.~(\ref{Eq:delta-pot-neut})--(\ref{Eq:delta-pot-top}).

\subsection{Different contributions}
\label{subsec:different-contributions}

For fixed values of $\alpha_2$ and $\alpha_3$, as in Fig.~\ref{Fig:h_j11-0},
we show in Fig.~\ref{Fig:h_j11-masks} the ratio of the one-loop corrected
trilinear Higgs coupling $\lambda_{111}$ and the corresponding tree-level
coupling.  This ratio is highly dependent on $\alpha_1$ (as is the coupling
itself).

\begin{figure}[htb]
\begin{center}
\includegraphics[width=95mm]{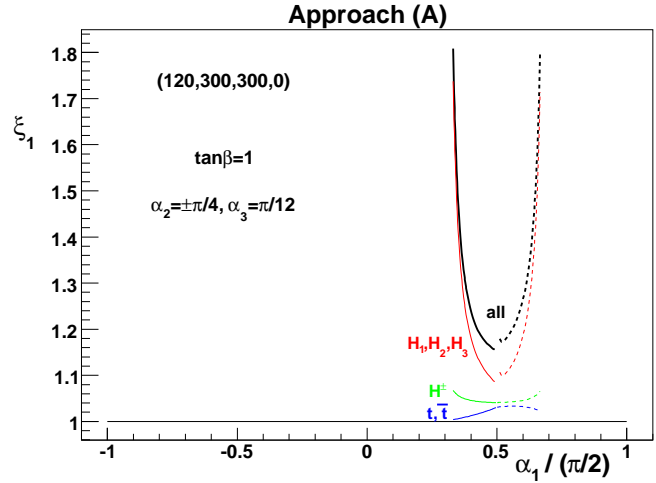}
\vspace*{-2mm}
\caption{\label{Fig:h_j11-masks} Contributions to the trilinear coupling ratio
$\xi_1$ plotted as functions of $\alpha_1$, for $\alpha_2=\pm\pi/4$, both with
$\alpha_3=\pi/12$.  The Higgs mass and parameters are as in
Fig.~\ref{Fig:h_j11-0}.  The scale parameter is $Q =2M_1^\text{LO}$.  The full
correction is shown by the upper, heavy, black lines, whereas partial
contributions from neutral and charged Higgs bosons, as well as from top
quarks, are shown by the thin, red, green and blue lines.}
\end{center}
\end{figure}

As compared with the one-loop corrections to the masses, the corrections to
the trilinear couplings are relatively much larger.  The following two points
are worth noting: (i) Most of the one-loop contribution comes from the
neutral-Higgs term, Eq.~(\ref{Eq:delta-pot-neut}) (red lines labeled
``$H_1,H_2,H_3$'').  (ii) The correction can be very large at the boundary of
the physically allowed region (in this case, in $\alpha_1$).  These regions of
very large corrections, which are excluded from averages over $\vecalpha$,
will be discussed further in the following.

In approach (A), when we scan over the values of $\vecalpha$, there are
regions where $M_3$ gets very close to $M_2$. From Ref.~\cite{Khater:2003wq},
we see that (at tree level) this happens when (i)~$\alpha_2 \to 0$ or
(ii)~$\tan\beta\to\cot\alpha_1$, or (iii)~$\alpha_2\to\pm\pi/2$.  Case (ii)
occurs for $\tan\beta=1$ and $\alpha_1=\pi/4$.  In this region, the trilinear
coupling tends to get very large. This is clearly a region of parameter space
where the present approach breaks down. In the notation of
Appendix~\ref{App:cubic} [see Eq.~(\ref{Eq:solutions})], we see that two roots
merge ($M_3\to M_2$) when the complex phases of $s_1$ and $s_2$ differ by
$2\pi/3$. Near such points, the derivatives of $s_1$ and $s_2$ with respect to
$a$ and $b$, particularly the second and third-order ones, tend to get very
large.  Actually, such points have been removed from the following figures,
and we impose a cut when scanning over $\vecalpha$ in
Sec.~\ref{subsect:averages}:
\begin{equation}
\min\{(M_3^\text{LO}-M_2^\text{LO}),(M_3-M_2)\}>5~\text{GeV}.
\end{equation}

\subsection{Scale dependence}

Just as in the case of one-loop corrections to the masses, we note a strong
dependence of the trilinear couplings on the scale parameter that enters 
the effective potential.
This scale plays a role somewhat analogous to the mass scale of superpartners
in the effective-potential approach to the MSSM.

\begin{figure}[hbt]
\begin{center}
\includegraphics[width=95mm]{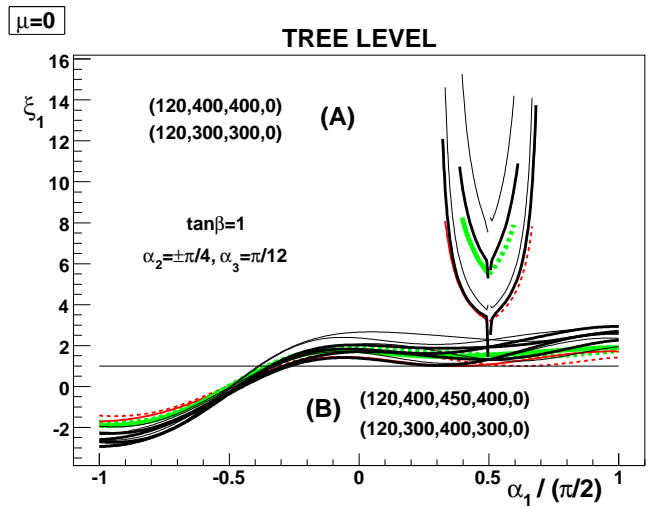}
\vspace*{-2mm}
\caption{\label{Fig:h_j11} Trilinear coupling ratio $\xi_1$ [see
Eq.~(\ref{Eq:xi-def})] as a function of $\alpha_1$.  
Colored lines refer to tree-level results (same color coding 
and parameters as in Fig.~\ref{Fig:h_j11-0}),
black lines refer to loop-corrected results (thin: $Q=2M_1^\text{LO}$,
heavy: $Q=500~\text{GeV}$).}
\end{center}
\end{figure}

In Fig.~\ref{Fig:h_j11} we study the one-loop corrected trilinear couplings,
for both the approaches (A) and (B), and for two sets of Higgs masses within
each approach (the same as in Fig.~\ref{Fig:h_j11-0}). 
We show how the one-loop corrections depend on the scale
$Q$. For this purpose we have considered two timelike values of $Q$,
namely, $Q=2M_1^\text{LO}$ and $Q=500~\text{GeV}$. As in the case of
Fig.~\ref{Fig:h_j11-masks}, we see that in approach (A), the couplings tend to
get large near the boundary of the allowed parameter space.
(Near $\alpha_1=\pi/4$, we notice the irregular behavior discussed in 
Sec.~\ref{subsec:different-contributions}, when $M_3\to M_2$.)

The dependence on the scale is rather involved, but there is a tendency that a
lower scale gives a larger correction.  This is rather clear in the upper part
of the figure, which is a plot for approach (A). For approach (B), however, we
have not labeled the individual curves, the main point being to show that
there is a considerable uncertainty related to the choice of scale in both
approaches.

\subsection{Decoupling}
\label{SubSec:decoupling}

In the context of the trilinear Higgs coupling, decoupling would imply
that the self-coupling of the light Higgs boson could be expressed
in terms of its mass, as in the Standard Model.
This was indeed found to hold for the MSSM, if the mass is taken to be
the loop-corrected one \cite{Hollik:2001px,Dobado:2002jz}.

\begin{figure}[hbt]
\begin{center}
\includegraphics[width=95mm]{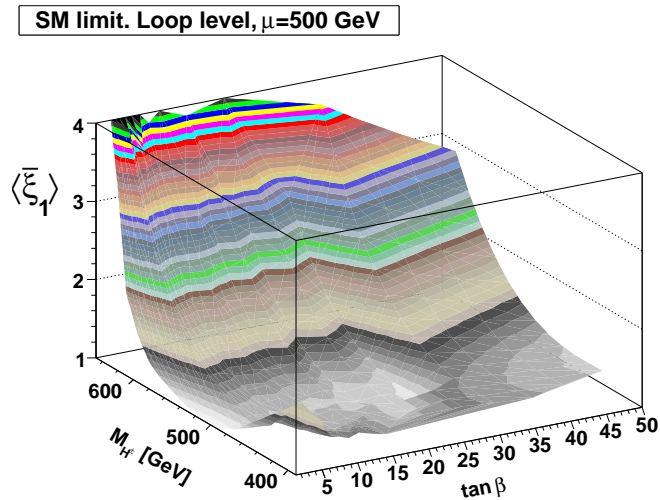}
\vspace*{-2mm}
\caption{\label{Fig:xi-sm} Trilinear coupling ratio $\langle\bar\xi_1\rangle$
as defined by Eq.~(\ref{Eq:bar-xi-1}). Different quantities in 
approach (A) with
$M_1^\text{LO}=120~\text{GeV}$, $M_2^\text{LO}=\mu=500~\text{GeV}$, and
$Q=2M_1^\text{LO}$.  The ratio $\langle\bar\xi_1\rangle\simeq1$ demonstrates
approximate decoupling for $M_2=\mu\sim M_{H^\pm}$.}
\end{center}
\end{figure}

In order to discuss decoupling in the 2HDM, we first need to define, as a
reference, a Standard-Model-like limit by imposing the tree-level constraint
(\ref{Eq:lambda6=lambda7=0}), together with Eq.~(\ref{Eq:no-cpv}). The SM-like
nature of this limit may be extended to the loop level, if in the loop
corrections (\ref{Eq:V-loop}) we include only the contributions of the
lightest neutral Higgs boson $H_1$ and the top quark.  Within this framework,
we may then study decoupling, by defining the ratio, analogous to
(\ref{Eq:xi-def-avg}):
\begin{equation} \label{Eq:bar-xi-1}
\langle\bar\xi_1\rangle
=\frac{\langle\lambda_{111}^\text{full}\rangle}
{\langle\lambda_{111}^\text{SM}\rangle},
\end{equation}
where
\begin{itemize}
\item
$\lambda_{111}^\text{SM}$ includes only loop corrections due to $t\bar t$ and
$H_1$, as would be the case in the SM.
\item
$\lambda_{111}^\text{full}$ includes all one-loop corrections to the 
Higgs coupling in the two Higgs doublet model, i.e., also
those due to $H_2$, $H_3$ and $H^{\pm}$.
\item
The averaging over $\alpha_2$ and $\alpha_3$ is constrained by
Eq.~(\ref{Eq:no-cpv}) with $\alpha_0=0.025\times\pi/2$, 
whereas $\alpha_1$ is left unconstrained.
\end{itemize}

It should be noted that here we do not insist on a correspondence between
the 2HDM and the MSSM, since this would require also $\alpha_1\simeq\beta$.

If the ratio (\ref{Eq:bar-xi-1}) were to come out as unity, it would be a
signal for decoupling: the heavy Higgs bosons do not affect the coupling of
the lightest one with itself. In Fig.~\ref{Fig:xi-sm}, we display this ratio,
for rather high values of $M_2^\text{LO}$ and $\mu$:
$M_2^\text{LO}=\mu=500~\text{GeV}$, subjecting all parameters to the general
constraints discussed in Sec.~\ref{subsect:tree-level-couplings-neutral}.  For
$M_{H^\pm}\sim M_2$, the ratio (\ref{Eq:bar-xi-1}) is indeed rather close to
unity, {\it but for larger values of $M_{H^\pm}$, it becomes significantly
larger}.  In the region where $\langle\bar\xi_1\rangle\lsim1.20$, we have also
checked that $M_3-M_2<{\cal O}(40)~\text{GeV}$. Thus, in this region of the
parameter space, decoupling holds at the loop level to about 10--20\%.  For
larger values of $M_{H^\pm}$, however, where the model is still consistent,
decoupling is strongly violated.


\begin{figure}[htb]
\begin{center}
\includegraphics[width=150mm]{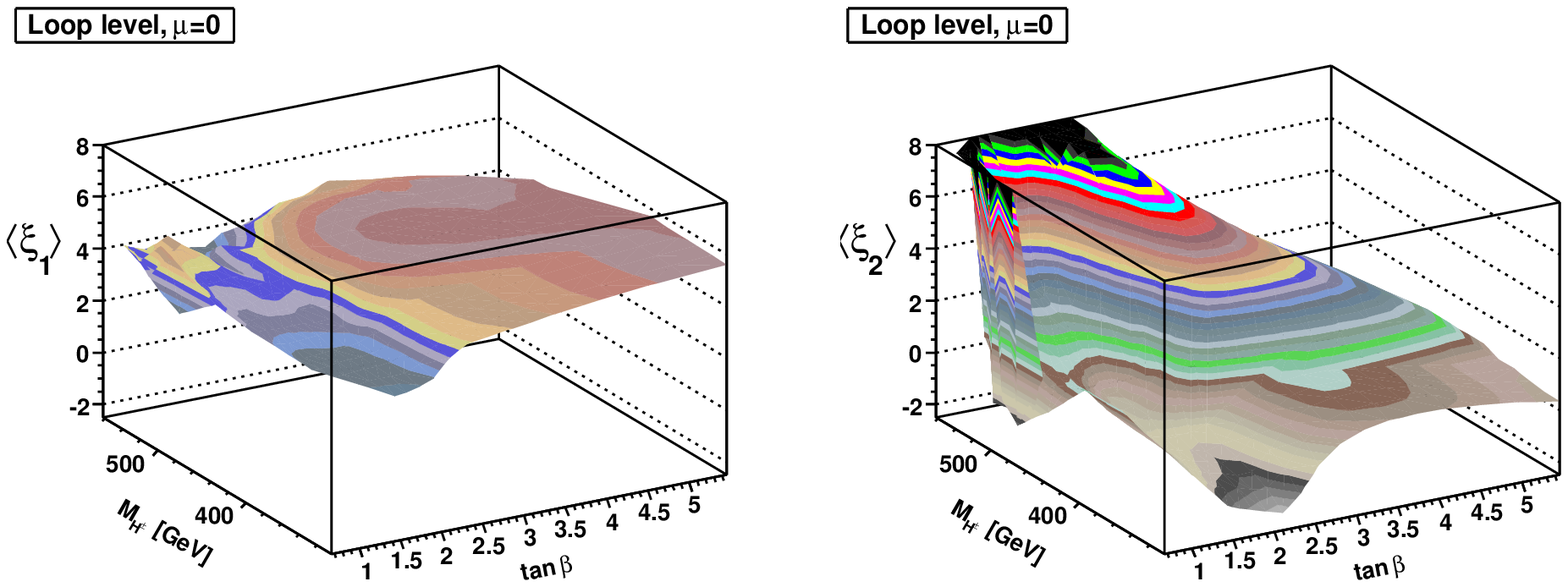}
\includegraphics[width=150mm]{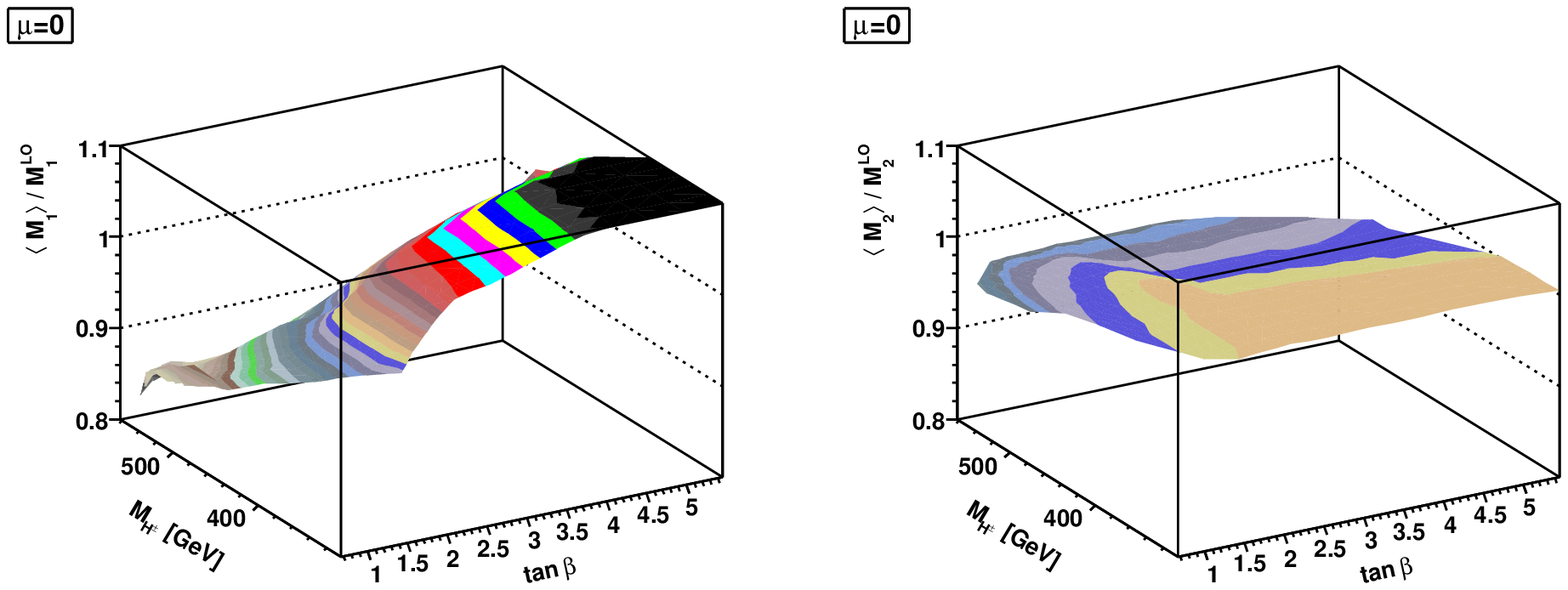}
\includegraphics[width=150mm]{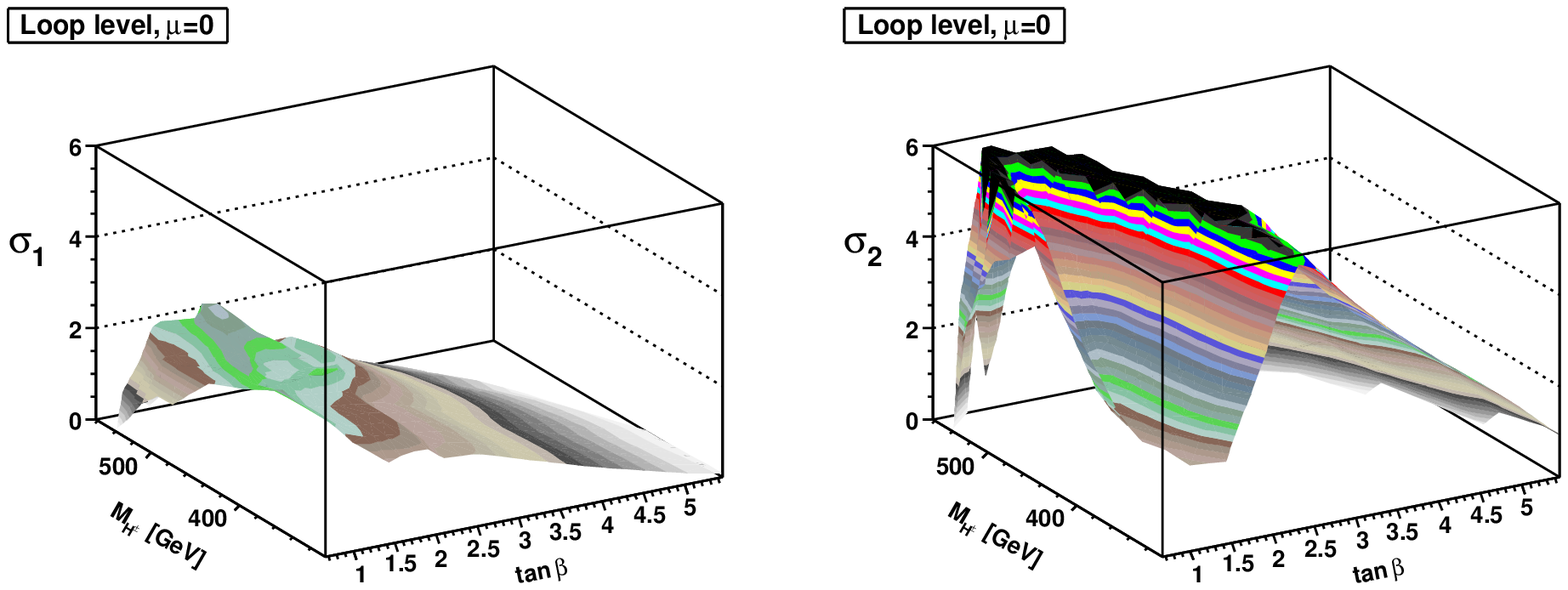}
\vspace*{-5mm}
\caption{\label{Fig:h_j11-1loop-level-mu=0} Approach (A) with
$M_1^\text{LO}=120~\text{GeV}$, $M_2^\text{LO}=300~\text{GeV}$, $\mu=0$ and
scale $Q=2M_1^\text{LO}$.  Top panel: Trilinear coupling ratios
$\langle\xi_1\rangle$ and $\langle\xi_2\rangle$ [see
Eq.~(\ref{Eq:xi-def-avg})].  Middle: one-loop multiplicative mass correction.
Bottom: Variance of trilinear coupling ratios $\xi_1$ and $\xi_2$.}
\end{center}
\end{figure}
\end{widetext}

The result displayed in Fig.~\ref{Fig:xi-sm} is quite stable
under a change of $\alpha_0$ from $0.025\times\pi/2$ to $0.05\times\pi/2$.
However, the result depends on the choice of scale adopted for
evaluating the one-loop potential. For example, for $Q=500~\text{GeV}$, 
$\langle\bar\xi_1\rangle$ differs significantly from unity. This is perhaps 
contrary to expectations, since $\log(M_2^\text{LO}/Q)$ then vanishes. 
However, we recall that (i) the one-loop potential also contains 
non-logarithmic terms, and (ii) the one-loop potential modifies the 
minimization conditions, i.e.,
its contribution shifts the values of the soft parameters 
$m_{11}^2$ and $m_{22}^2$ of the tree level potential.

In the 2HDM, in contrast to the MSSM,
the decoupling cannot be exact. The reason is obvious:
the heavy fields that we would like to neglect, are all bosonic,
and thus all contribute with the same sign to the potential.
Furthermore, even if we choose the scale $Q$ such that the logarithms
vanish, there would still be the non-logarithmic remainders.

The question of decoupling can be illustrated by the behavior of the $\rho$
parameter, which gets contributions from all {\it pairs} of Higgs bosons.
These contributions vanish for equal masses, but are large when the masses are
very different.

In the limit of no CP violation, custodial symmetry \cite{Weinberg:1972fn}
would imply $M_{H^\pm}=M_j$, where $M_j$ is either the mass of the CP odd
($A$) or a CP even ($H$) Higgs particle \cite{Gerard:2007kn}.  In the case
that $M_{H^\pm}=A$, orthogonality of the even-sector mixing matrix would
protect the $\Delta\rho$ parameter from large corrections. In the present case
of CP violation, a higher degeneracy is required, $M_2=M_3=M_{H^\pm}$, in
order to avoid large contributions to $\Delta\rho$.  Since this is in general
not satisfied, the experimental constraints on $\Delta\rho$ severely constrain
the allowed parameter space \cite{WahabElKaffas:2007xd}, and there is no
decoupling.

\begin{widetext}

\begin{figure}[htb]
\begin{center}
\includegraphics[width=150mm]{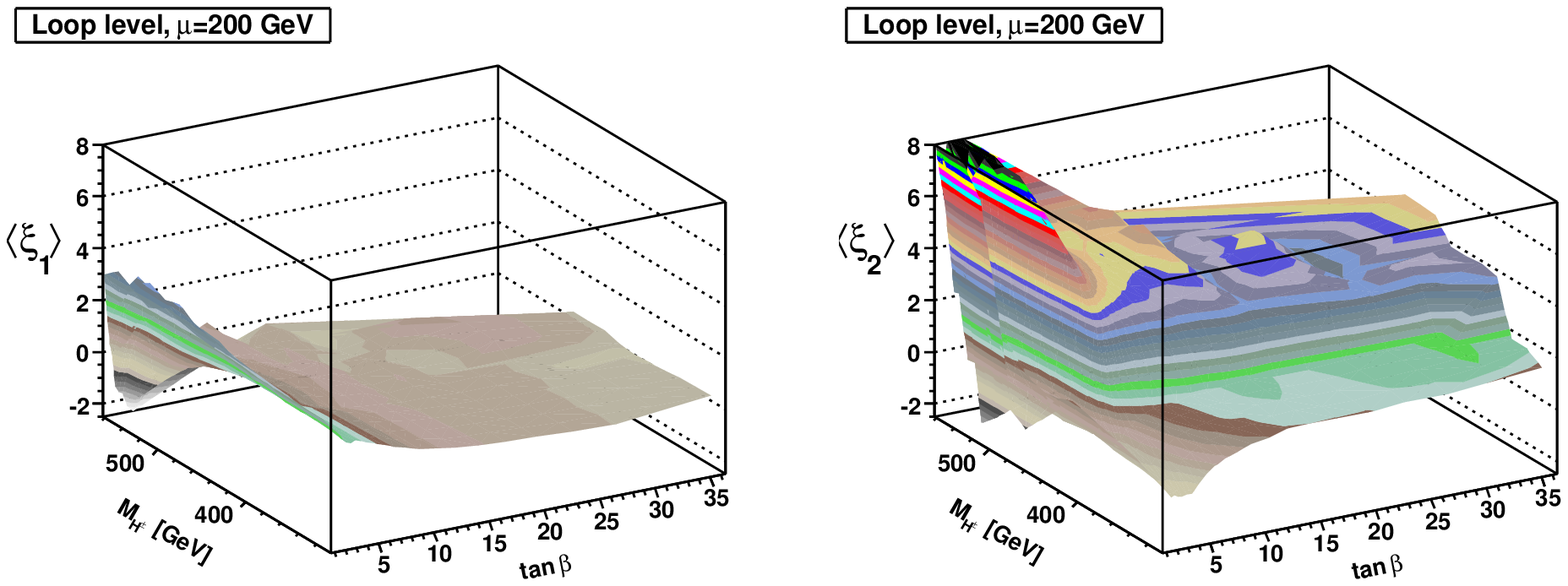}
\includegraphics[width=150mm]{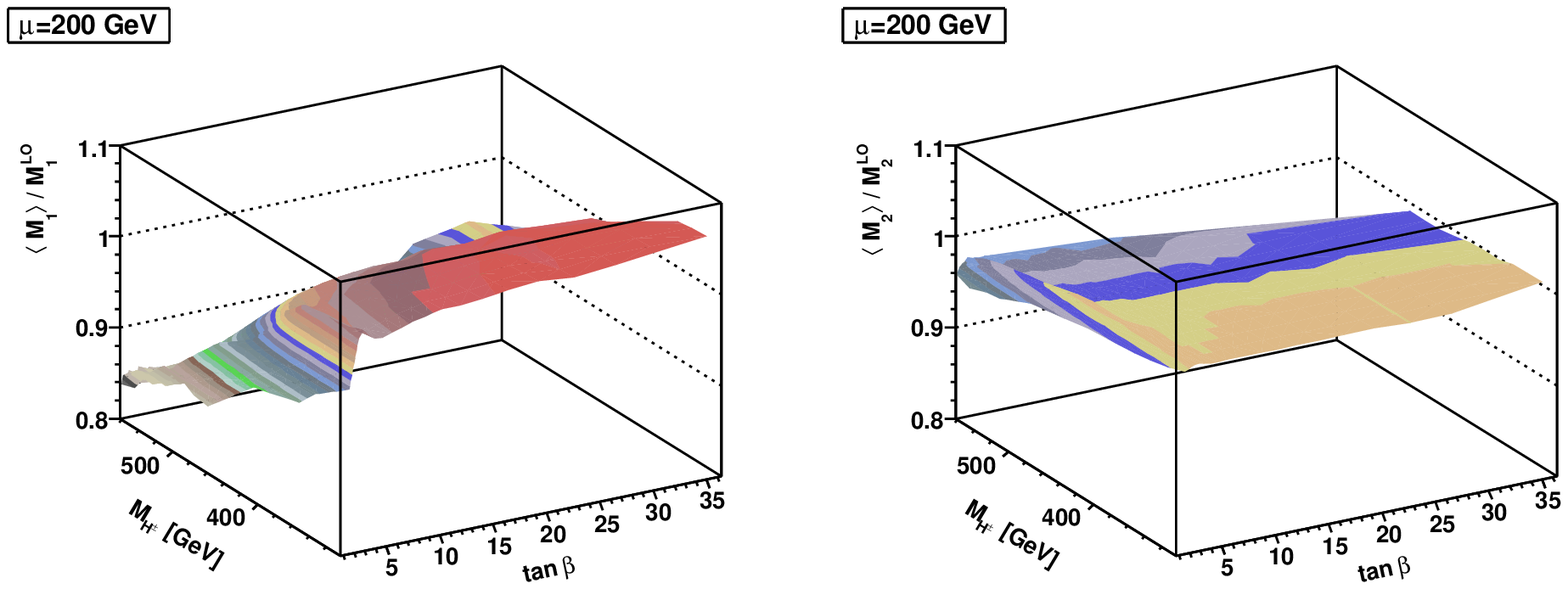}
\includegraphics[width=150mm]{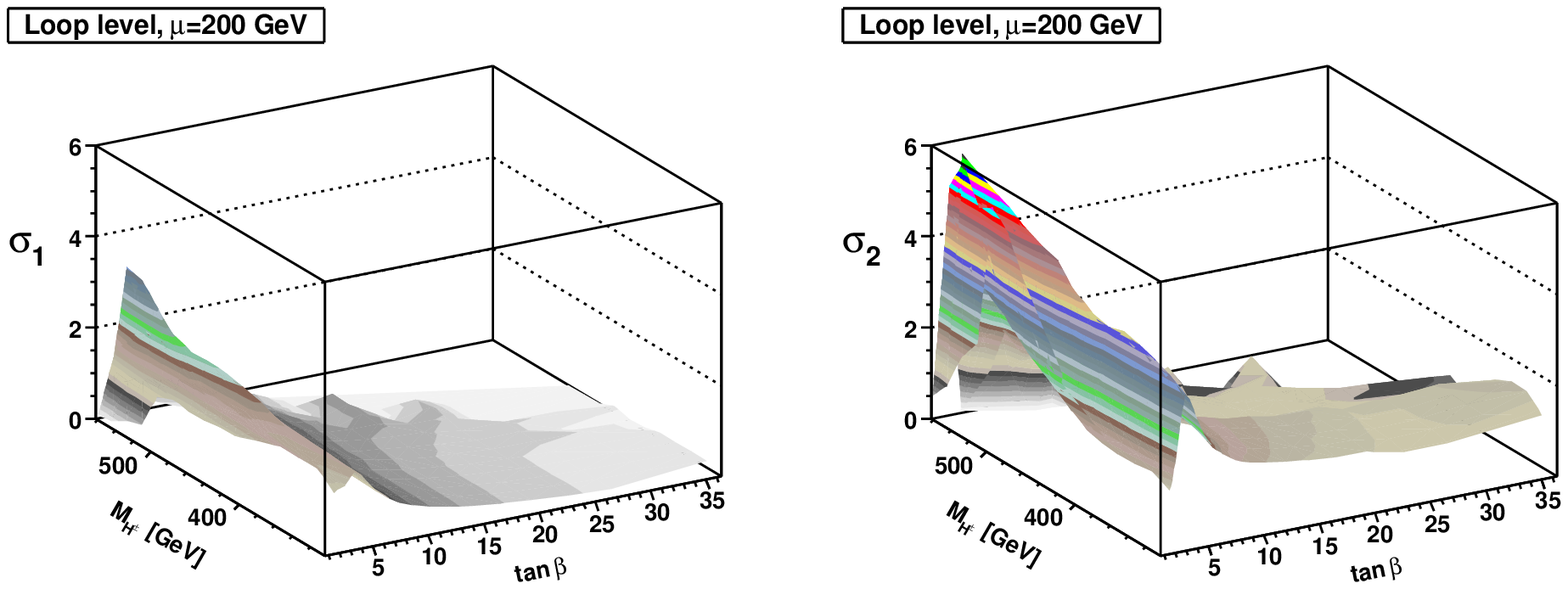}
\vspace*{-5mm}
\caption{\label{Fig:h_j11-1loop-level-mu=200}  Same as
Fig.~\ref{Fig:h_j11-1loop-level-mu=0}, but with $\mu=200~\text{GeV}$.}
\end{center}
\end{figure}
\end{widetext}

\subsection{The general case}
\label{subsect:averages}

After this exploratory discussion, we are now ready to scan and average over
the full range of $\vecalpha$, compatible with theoretical and experimental
constraints on the model, as was done at tree level in
Sec.~\ref{subsect:tree-level-couplings-neutral}.
Figs.~\ref{Fig:h_j11-1loop-level-mu=0} and \ref{Fig:h_j11-1loop-level-mu=200}
are plotted for approach (A) with $M_1^\text{LO}=120~\text{GeV}$,
$M_2^\text{LO}=300~\text{GeV}$, $\mu=0$
(Fig.~\ref{Fig:h_j11-1loop-level-mu=0}), $\mu=200~\text{GeV}$
(Fig.~\ref{Fig:h_j11-1loop-level-mu=200}) and with scale $Q=2M_1^\text{LO}$ in
both cases.

In these figures, we show the ratios $\langle\xi_1\rangle$ 
and $\langle\xi_2\rangle$ of trilinear couplings in the top panels.
The middle panels are devoted to the loop-induced corrections
to the masses $M_1$ and $M_2$, whereas the bottom panels
show the variance associated with the averaging of $\xi_1$
and $\xi_2$ over the $\vecalpha$ parameter space.

An important difference between the two cases presented in these figures is
the range in $\tan\beta$. As already discussed, for a low value of $\mu$, the
high values of $\tan\beta$ would lead to a model that would violate unitarity,
and are, therefore, excluded \cite{ElKaffas:2006nt}.

Comparing the top panels of Figs.~\ref{Fig:h_j11-1loop-level-mu=0}
and \ref{Fig:h_j11-1loop-level-mu=200} with the corresponding
tree-level resuls shown in Fig.~\ref{Fig:h_j11-tree-level}, we note the
following: (i) The overall shape in the $\tan\beta$--$M_{H^\pm}$ plane is
similar.  (ii) The loop-corrected couplings are somewhat larger, after
averaging over $\boldsymbol{\alpha}$, in analogy with the results shown in
Fig.~\ref{Fig:h_j11} for $\alpha_2$ and $\alpha_3$ fixed.  

The center panel of Fig.~\ref{Fig:h_j11-1loop-level-mu=0} is rather similar to
Fig.~\ref{Fig:mass-corr}, the only difference being that here the LEP2
constraint is imposed at the loop level.  Focusing first on the loop
corrections to $M_1$, and comparing the mass correction at the two different
values of $\mu$, we note that the case $\mu=0$ gives a larger reduction at low
values of $\tan\beta$ and a larger increase at high values of $\tan\beta$.
For $M_2$, on the other hand, the loop correction has only a rather weak
dependence on $\tan\beta$ (and $M_{H^\pm}$).
For a higher value of the scale $Q$, however, the loop correction to $M_1$
may be positive also at low values of $\tan\beta$.

The trilinear couplings represented in these figures, are of course not
physical, they are instead ``typical values'', obtained by an averaging over
the angles of the mixing matrix $R$ that diagonalizes the mass-squared matrix.
As discussed above, the variation over the $\vecalpha$ parameter space can be
considerable, as illustrated by the {\it variance} shown in the bottom panels.

A more detailed comparison reveals that the loop corrections are most
important for low values of low-$\tan\beta$. This is particularly so
in the case of  $\langle\xi_2\rangle$, where the variance is also quite 
large.  This is in part caused by the ill-defined limit 
$M_3\to M_2$ ($\alpha_2\to0$) discussed above. [Actually, at high values of 
$M_{H^\pm}$, the $\langle\xi_2\rangle$ plots are truncated at the top.]

\begin{widetext}

\begin{figure}[htb]
\begin{center}
\includegraphics[width=150mm]{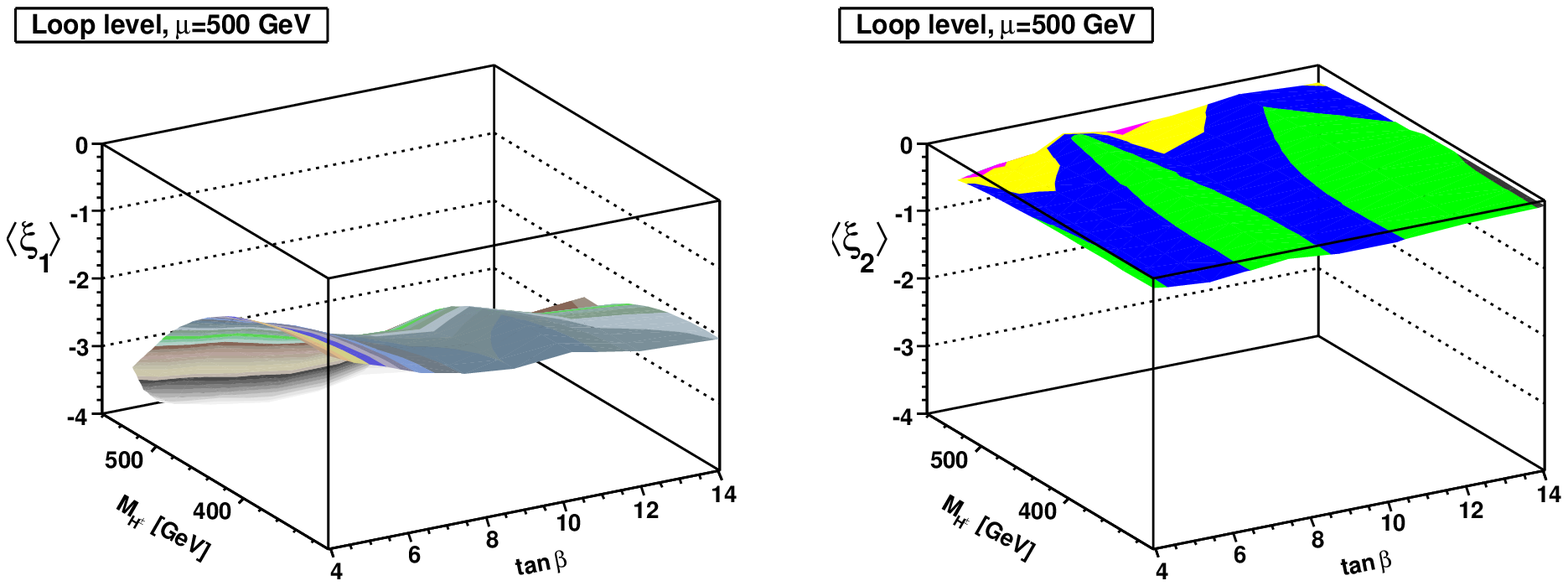}
\includegraphics[width=150mm]{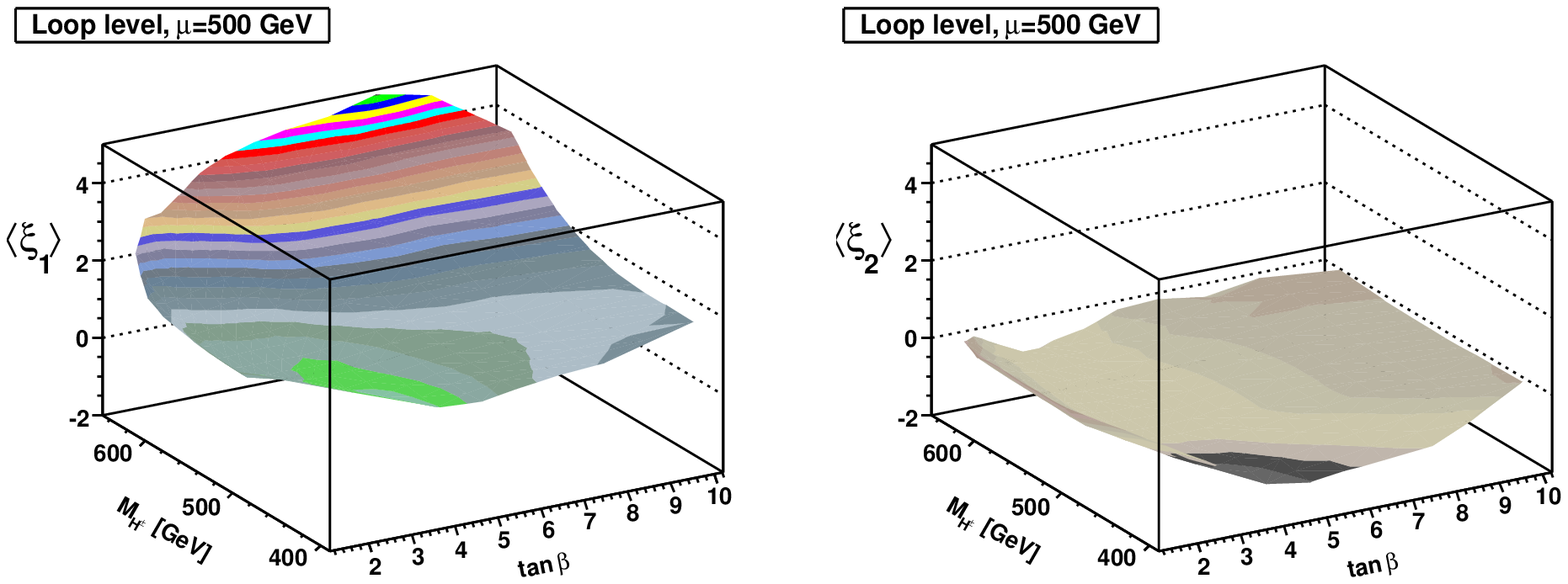}
\vspace*{-5mm}
\caption{\label{Fig:h_j11-1loop-level-mu=500} Similar to 
the upper panels of Figs.~\ref{Fig:h_j11-1loop-level-mu=0}
and \ref{Fig:h_j11-1loop-level-mu=200}, but for 
$\mu=500~\text{GeV}$.
Upper panels: $M_2^\text{LO}=300~\text{GeV}$;
lower panels: $M_2^\text{LO}=500~\text{GeV}$.
Note that the scales are different from those in the earlier figures.}
\end{center}
\end{figure}
\end{widetext}

The correlation between these trilinear couplings and the quartic couplings
$\lambda_i$ of the tree-level potential or their averages,
$\langle\lambda_i\rangle$, is not very simple, but a couple of observations
can be made.  In Fig.~\ref{Fig:h_j11-1loop-level-mu=0}, for $\mu = 0$ the
enhancement at low values of $\tan\beta$ is correlated with a corresponding
enhancement of $\langle\lambda_2\rangle$, whereas $\langle\lambda_1\rangle$ is
large at high values of $\tan\beta$.

For the $H_1H_1H_1$ coupling, represented by $\langle\xi_1\rangle$, we note
that for $\mu=0$ it is rather large for all values of $\tan\beta$ and
$M_{H^\pm}$ where the model is consistent, whereas for $\mu=200~\text{GeV}$ it
falls to zero for large values of $\tan\beta$.

Finally, in Fig.~\ref{Fig:h_j11-1loop-level-mu=500} we show the behavior for
somewhat higher value of the soft mass parameter, $\mu=500~\text{GeV}$.  We
display the averages $\langle\xi_1\rangle$ and $\langle\xi_2\rangle$ for two
values of $M_2^\text{LO}$, 300~GeV (as was adopted in the earlier plots) and
500~GeV.  Compared to the corresponding panels of
Figs.~\ref{Fig:h_j11-1loop-level-mu=0} and \ref{Fig:h_j11-1loop-level-mu=200},
the present ones (for $\mu=500~\text{GeV}$) are more smooth, and also somewhat
different in magnitude.

\section{Summary and Conclusions}
\label{sect:Summary}
The two Higgs doublet model is a viable extension of the Standard Model
Higgs sector. Apart from having a richer structure of Higgs bosons and 
Higgs couplings, it can  accommodate CP violation beyond that of  
the Kobayashi--Maskawa mechanism in the SM. In this paper we have studied 
in detail the trilinear couplings of the lightest Higgs boson of this model,
including the CP violating effects. Our main objective has been to study
the implications of CP violation for the trilinear couplings of the
general two Higgs doublet model. We have discussed two different
parametrizations of the general two Higgs doublet model, and imposed
the constraints on the parameters of the model following from  
theoretical considerations and from different 
experiments. Tree-level unitarity, $B$ physics and $\Delta\rho$
are the most constraining ones for the parameters of the two 
Higgs doublet moddel with CP violation.

Within the allowed domain of the parameter space of the model, the trilinear
couplings can have a very strong dependence on the neutral Higgs boson mixing
angles.  Because of this strong dependence of the trilinear couplings on the
mixing angles, we have presented averages over the allowed ranges for these
angles. Together with the variances, these show the range of trilinear
couplings that can be expected in the two Higgs doublet model with CP
violation.

There is a region of parameter space of the 2HDM, which we call ``{\it minimal
CP violation},'' which corresponds to the domain of parameters close to the
Higgs sector of the CP-conserving minimal supersymmetric standard model.  In
much of this parameter space there is a considerable enhancement of the
trilinear couplings of the two Higgs doublet model as compared with the
corresponding trilinear coupling of the SM.

We have computed the radiative corrections to the masses of the Higgs bosons
and the trilinear couplings in the one-loop effective potential approximation.
The one-loop corrected mass of the lightest neutral Higgs boson can be
modified by as much as 20\%, whereas the heavier ones are less affected
by the radiative corrections.  The
one-loop corrected trilinear couplings are typically enhanced by the loop
effects, but so is the sensitivity to the neutral-sector mixing angles.  A
similar enhancement due to CP violating effects has also been found for the
MSSM \cite{Dubinin:2002nx,Philippov:2006th}.

To a good approximation, decoupling holds in a limited range of parameters,
where the model is close to the CP-conserving MSSM. In general, however, the
heavier states also have a considerable impact  on the properties of the
lightest Higgs boson. In this sense there is no decoupling in the
two Higgs doublet model with CP violation.

\section{Acknowledgements}
The research of PO has been supported in part by  
the Research Council of Norway.
The work of PNP is supported by the Board of Research in Nuclear Sciences,
India under project No.~2007/37/34/BRNS/1970.  
He would like to thank the the Abdus Salam ICTP, Trieste, and 
Department of Physics and Technology, University of Bergen for hospitality 
where part of this work was done.  LS is grateful to
Department of Physics and Technology, University of Bergen for its
generous hospitality,  where part of this work was done, and The
Scientific and Technological Research Council of Turkey (TUBITAK)
for financial support. It is  a great pleasure for LS
to thank Zekeriya Aydin and Dumus Ali Demir for useful discussions
in the early stages of this work.

\appendix

\section{Quartic couplings expressed in terms of masses}
\label{App:quartic}
In the 2HDM, with $\Im\lambda_5\ne0$, the two masses $M_1$ and $M_2$,
together with $\vecalpha$ and $\tan\beta$, determine $M_3$.  The masses will
also be related to the quartic couplings of the Higgs potential,
via linear relations that can be unambiguously inverted.
It is convenient to distinguish two cases, whether or not $\lambda_6$
and $\lambda_7$ are non-zero.
\subsection{Approach (A), $\boldsymbol{\lambda_6=\lambda_7=0}$}
This case is referred to as approach (A) in the text.  Providing also
 $M_{H^\pm}$ and $\mu^2$, in addition to $M_1$, $M_2$ and $\vecalpha$, all the
 $\lambda$'s are determined as follows \cite{ElKaffas:2006nt,ElKaffas:2007rq}:
\begin{align} \label{Eq:lambda1}
\lambda_1&=\frac{1}{c_\beta^2v^2}
[c_1^2c_2^2M_1^2
+(c_1s_2s_3+s_1c_3)^2M_2^2 \nonumber \\
&+(c_1s_2c_3-s_1s_3)^2M_3^2
-s_\beta^2\mu^2], \\
\lambda_2&=\frac{1}{s_\beta^2v^2} \label{Eq:lambda2}
[s_1^2c_2^2M_1^2
+(c_1c_3-s_1s_2s_3)^2M_2^2 \nonumber \\
&+(c_1s_3+s_1s_2c_3)^2M_3^2-c_\beta^2\mu^2], \\
\lambda_3&=\frac{1}{c_\beta s_\beta v^2} \label{Eq:lambda3}
\{c_1s_1[c_2^2M_1^2+(s_2^2s_3^2-c_3^2)M_2^2 \nonumber \\
&+(s_2^2c_3^2-s_3^2)M_3^2]
+s_2c_3s_3(c_1^2-s_1^2)(M_3^2-M_2^2)\} \nonumber \\
+&\frac{1}{v^2}[2M_{H^\pm}^2-\mu^2], \\
\lambda_4&=\frac{1}{v^2}
[s_2^2M_1^2+c_2^2s_3^2M_2^2+c_2^2c_3^2M_3^2
+\mu^2-2M_{H^\pm}^2], \\
\Re\lambda_5&=\frac{1}{v^2}
[-s_2^2M_1^2-c_2^2s_3^2M_2^2-c_2^2c_3^2M_3^2+\mu^2], \label{Eq:lambda5R} \\
\Im\lambda_5&=\frac{-1}{c_\beta s_\beta v^2}
\{c_\beta[c_1c_2s_2M_1^2-c_2s_3(c_1s_2s_3+s_1c_3)M_2^2 \nonumber \\
&+c_2c_3(s_1s_3-c_1s_2c_3)M_3^2] 
+s_\beta[s_1c_2s_2M_1^2 \label{Eq:lambda5I} \\
&+c_2s_3(c_1c_3\!-\!s_1s_2s_3)M_2^2
\!-\!c_2c_3(c_1s_3\!+\!s_1s_2c_3)M_3^2]\}, \nonumber 
\end{align}
where $c_\beta=\cos\beta$, $s_\beta=\sin\beta$.

While $M_3^2$ is given in terms of $M_1^2$, $M_2^2$, $R$ and $\tan\beta$, it
is more transparent not to substitute for $M_3^2$ in the expressions
(\ref{Eq:lambda1})--(\ref{Eq:lambda5I}).  These equations are the analogues of
those of \cite{Casalbuoni:1987cz} for the CP-conserving 2HDM.
\subsection{Approach (B), $\boldsymbol{\lambda_6\neq0}$, 
$\boldsymbol{\lambda_7\neq0}$}
In this case, referred to as approach (B) in the text, we take $M_1$, $M_2$,
$M_2$, $\vecalpha$, $M_{H^\pm}$ and $\mu^2$, together with $\Im\lambda_5$,
$\Re\lambda_6$ and $\Re\lambda_7$, as the input.  In order to keep the
notation compact, it is convenient to introduce the following abbreviations:
\begin{align}
\Re\lambda_{345}&=\lambda_3+\lambda_4+\Re\lambda_5, \\
\Re\lambda_{567}&=\Re\lambda_{5}
+\cot\beta\,\Re\lambda_6+\tan\beta\,\Re\lambda_7.
\end{align}
Then, the $\lambda$'s can be determined from the
following relations:
\begin{align}
\lambda_1&=\frac{1}{c_\beta^2}
\biggl[\frac{{\cal M}_{11}^2-s_\beta^2\mu^2}{v^2}
\!-\!\frac{s_\beta}{2c_\beta}
(3c_\beta^2\Re\lambda_6-s_\beta^2\Re\lambda_7)\biggr], \\
\lambda_2&=\frac{1}{s_\beta^2} 
\biggl[\frac{{\cal M}_{22}^2-c_\beta^2\mu^2}{v^2}
\!+\!\frac{c_\beta}{2s_\beta}
(c_\beta^2\Re\lambda_6-3s_\beta^2\Re\lambda_7)\biggr], \\
\Re\lambda_{345}&=\frac{1}{c_\beta s_\beta}
\biggl[\frac{c_\beta s_\beta\mu^2+{\cal M}_{12}^2}{v^2} \nonumber \\
&-\frac{3}{2}
(c_\beta^2\Re\lambda_6+s_\beta^2\Re\lambda_7)\biggr], \\
\lambda_4&=\frac{2}{v^2}
[\mu^2-M_{H^\pm}^2]-\Re\lambda_{567}, \\
\Re\lambda_5&=\frac{\mu^2-{\cal M}_{33}^2}{v^2} 
-\frac{1}{2c_\beta s_\beta}
(c_\beta^2\Re\lambda_6+s_\beta^2\Re\lambda_7), \\
\Im\lambda_6&=-\frac{1}{2c_\beta}\biggl[\frac{2{\cal M}_{13}^2}{v^2}
+s_\beta\Im\lambda_5\biggr], \\
\Im\lambda_7&=-\frac{1}{2s_\beta}\biggl[\frac{2{\cal M}_{23}^2}{v^2}
+c_\beta\Im\lambda_5\biggr].
\end{align}
Invoking Eq.~(\ref{Eq:calM-RMsqR}), the ${\cal M}_{ij}^2$ can be
expressed in terms of $R$ and the masses $M_1$, $M_2$ and $M_3$.
\section{Trilinear coupling for some special cases}
\label{App:trilin_111_112}
In order to get a feeling for the trilinear Higgs couplings of the 2HDM, 
we shall here explicitly write down the two trilinear couplings 
$\lambda_{111}$ and $\lambda_{112}$ for some special cases.
\subsection{The trilinear coupling $\boldsymbol{\lambda_{111}}$}
We start by writing down Eq.~(\ref{Eq:neutral-3-coupl}) explicitly
for $\{i, j, k\}=\{1, 1, 1\}$:
\begin{align}
\lambda_{111}
&=\sum_{m\le n\le o=1,2,3}^\ast R_{1m}R_{1n}R_{1o}\,a_{mno}  \nonumber \\
&=3!\{R_{11}^2[R_{11}\,a_{111}+R_{12}\,a_{112}+R_{13}\,a_{113}] \nonumber \\
&+R_{11}[R_{12}^2a_{122}+R_{12}R_{13}\,a_{123}+R_{13}^2\,a_{133}] \nonumber \\
&+R_{12}[R_{12}^2\,a_{222}+R_{12}R_{13}\,a_{223}+R_{13}^2\,a_{233}]
\nonumber \\
&+R_{13}^3\,a_{333}\},
\end{align}
where the factor $3!$ is due to the fact that we here have couplings to three
identical $H_1$ fields.  Substituting for the elements of the rotation
matrix, Eq.~(\ref{Eq:R-angles}), one obtains
\begin{align}
\lambda_{111}
&=3!\{(c_1c_2)^2[c_1c_2\,a_{111}+s_1c_2\,a_{112}+s_2\,a_{113}] \nonumber \\
&+c_1c_2[(s_1c_2)^2\,a_{122}+s_1c_2s_2\,a_{123}+s_2^2\,a_{133}] \nonumber \\
&+s_1c_2[(s_1c_2)^2\,a_{222}+s_1c_2s_2\,a_{223}+s_2^2\,a_{233}] \nonumber \\
&+s_2^3\,a_{333}\}.
\end{align}

Let us now consider the case of $\lambda_6=\lambda_7=0$, and substitute
for the coefficients $a_{mno}$ from Eq.~(\ref{Eq:trilin-coupl-a_mno}).
In this case,
\begin{align} \label{Eq:lambda_111_explicit}
\lambda_{111}
&=3\{c_1c_2c_\beta(c_1^2c_2^2+s_2^2s_\beta^2)\lambda_1
+s_1c_2s_\beta(s_1^2c_2^2+s_2^2c_\beta^2)\lambda_2 \nonumber \\
&-2c_2s_2^2(c_1c_\beta+s_1s_\beta)\Re\lambda_{345}\nonumber \\
&-s_2[(c_2^2-s_2^2)c_\beta s_\beta+2c_1s_1c_2^2]\Im\lambda_5\}.
\end{align}
In the limit of no CP violation, with $H_1$ being odd, we have $c_2=0$
[see Eq.~(\ref{Eq:H_1-odd})],
and the trilinear coupling simplifies further to
\begin{equation}
\lambda_{111}=3s_2^3\,c_\beta s_\beta\Im\lambda_5,
\end{equation}
with $s_2=\pm1$. 
Similarly, in the limit of no CP violation, with $H_3$ being odd,
we have $s_2=s_3=0$, and
obtain the simple expression given in Eq.~(\ref{Eq:lambda:special-case-111}),
whereas the limit of no CP violation with $H_2$ being odd, leads to no further
simplification of the expression (\ref{Eq:lambda_111_explicit}).

\subsection{The trilinear coupling $\boldsymbol{\lambda_{112}}$}
We shall again first write  Eq.~(\ref{Eq:neutral-3-coupl}) explicitly
for $\{i, j, k\}=\{1, 1, 2\}$:
\begin{align}
\lambda_{112}
&=\sum_{m\le n\le o=1,2,3}^\ast R_{i^\prime m}R_{j^\prime n}R_{k^\prime o}\,
a_{mno}  \nonumber \\
&=2\sum_{m\le n\le o=1,2,3} 
\{R_{1m}R_{1n}R_{2o}+R_{1m}R_{2n}R_{1o} \nonumber\\
&+R_{2m}R_{1n}R_{1o}\}a_{mno} \nonumber\\
&=2\{3R_{11}^2R_{21}\,a_{111}+R_{11}(R_{11}R_{22}+2R_{12}R_{21})a_{112}
\nonumber \\
&+R_{11}(R_{11}R_{23}+2R_{13}R_{21})a_{113} \nonumber \\
&+R_{12}(2R_{11}R_{22}+R_{12}R_{21})a_{122} \nonumber \\
&+[R_{11}(R_{12}R_{23}+R_{13}R_{22})+R_{12}R_{13}R_{21}]a_{123} \nonumber \\
&+R_{13}(2R_{11}R_{23}+R_{13}R_{21})a_{133}
+3R_{12}^2R_{22}\,a_{222} \nonumber \\
&+R_{12}[R_{12}R_{23}+2R_{13}R_{22}]a_{223} \nonumber\\
&+R_{13}(2R_{12}R_{23}+R_{13}R_{22})a_{233}
+3R_{13}^2R_{23}\,a_{333}\}.
\end{align}
If we now substitute for the elements of the rotation matrix,
for the coefficients $a_{mno}$, and set $\lambda_6=\lambda_7=0$, 
we obtain the result quoted in Eq.~(\ref{Eq:lambda:special-case-112}).

\section{Derivatives of $M_\ell^2$}
\label{App:cubic}

The loop corrections to the trilinear couplings depend on derivatives of the
squared masses $M_\ell^2$ with respect to the weak fields $\eta_i$.  
In order to obtain these derivatives, a considerable amount of book-keeping 
is required. The complication arises from the fact that we allow for CP
non-conservation and $M_\ell^2$ will thus be determined by the roots of a
cubic equation rather than a quadratic one.  We start by considering the
$M_\ell^2$ as functions of the coefficients of the corresponding cubic
eigenvalue equation:
\begin{equation} \label{Eq:partialM_ell}
\frac{\partial M_\ell^2}{\partial \eta_i}
=\frac{\partial}{\partial \eta_i}\,M_\ell^2(a,b,c)
\end{equation}
where $M_\ell^2=\lambda_\ell$ is a solution of
\begin{equation} \label{Eq:cubic}
\lambda^3+a\lambda^2+b\lambda+c=0,
\end{equation}
with field-dependent coefficients:
\begin{equation}
a=a(\eta_1,\eta_2,\eta_3),\quad b=b(\eta_1,\eta_2,\eta_3), \quad
c=c(\eta_1,\eta_2,\eta_3).
\end{equation}
These coefficients of the cubic eigenvalue equation are obtained from the
derivatives (\ref{Eq:M_sq-def}), where, in contrast to Eq.~(\ref{Eq:M^2_ij}),
we {\it do not} set the fields to zero.
In terms of the $3\times3$ mass squared matrix ${\cal M}^2$, these coefficients
of the cubic equation are given by
\begin{align}
a&=-\Tr{\cal M}^2, \nonumber \\
b&=\half\bigl\{(\Tr{\cal M}^2)^2
-\Tr\bigl[({\cal M}^2)^2\bigr]\bigr\}, \nonumber \\
c&=-\det {\cal M}^2.
\end{align}

The derivatives (\ref{Eq:partialM_ell}) can thus be split up into simpler
entities:
\begin{align} \label{Eq:diff-Msq-eta-1}
\frac{\partial}{\partial \eta_i}\,M_\ell^2(a,b,c)
&=\frac{\partial M_\ell^2}{\partial a}\frac{\partial a}{\partial  \eta_i}
+\frac{\partial M_\ell^2}{\partial b}\frac{\partial b}{\partial  \eta_i}
+\frac{\partial M_\ell^2}{\partial c}\frac{\partial c}{\partial  \eta_i} 
\nonumber \\
&=\sum_\alpha \frac{\partial M_\ell^2}{\partial a_\alpha}
\frac{\partial a_\alpha}{\partial\eta_i},
\end{align}
where we have introduced the collective notation
\begin{equation}
a_\alpha=\{a,b,c\}.
\end{equation}

The higher derivatives can likewise be written as
\begin{align} \label{Eq:diff-Msq-eta-2}
\frac{\partial^2 M_\ell^2}{\partial\eta_i\partial\eta_j}
&=\sum_\alpha
\biggl[\sum_\beta
\frac{\partial^2 M_\ell^2}{\partial a_\alpha\partial a_\beta}
\frac{\partial a_\alpha}{\partial\eta_i}
\frac{\partial a_\beta}{\partial\eta_j}
+\frac{\partial M_\ell^2}{\partial a_\alpha}
\frac{\partial^2 a_\alpha}{\partial\eta_i\partial\eta_j}
\biggr], \\
\frac{\partial^3 M_\ell^2}{\partial\eta_i\partial\eta_j\partial\eta_k}
&=\sum_\alpha
\biggl\{\sum_\beta \biggl[
\sum_\gamma\frac{\partial^3M_\ell^2}
{\partial a_\alpha\partial a_\beta\partial a_\gamma}
\frac{\partial a_\alpha}{\partial\eta_i}
\frac{\partial a_\beta}{\partial\eta_j}
\frac{\partial a_\gamma}{\partial\eta_k} \nonumber \\
&\quad
+\frac{\partial^2 M_\ell^2}{\partial a_\alpha\partial a_\beta}
\biggl(
 \frac{\partial^2 a_\alpha}{\partial\eta_i\partial\eta_j}
 \frac{\partial a_\beta}{\partial\eta_k}
+\frac{\partial^2 a_\alpha}{\partial\eta_j\partial\eta_k}
 \frac{\partial a_\beta}{\partial\eta_i} \nonumber \\
&+\frac{\partial^2 a_\alpha}{\partial\eta_k\partial\eta_i}
 \frac{\partial a_\beta}{\partial\eta_j}\biggr) \biggr]
+\frac{\partial M_\ell^2}{\partial a_\alpha}
\frac{\partial^3 a_\alpha}{\partial\eta_i \partial\eta_j \partial\eta_k}
\biggr\}. \label{Eq:diff-Msq-eta-3}
\end{align}

\subsection{Cubic equation}

In order to obtain the derivatives of $M_\ell^2$ with respect to the
$a_\alpha$ that enter Eqs.~(\ref{Eq:diff-Msq-eta-1}),
(\ref{Eq:diff-Msq-eta-2}) and (\ref{Eq:diff-Msq-eta-3}), we start by solving
the cubic equation (\ref{Eq:cubic}) in terms of the following notation.  Let
\begin{equation} \label{Eq:q-r}
q=\frac{1}{3}\,b-\frac{1}{9}\,a^2, \quad
r=\frac{1}{6}(ab-3c)-\frac{1}{27}a^3
\end{equation}
and
\begin{equation}
s_1=[r+\Delta]^{1/3}, \quad
s_2=[r-\Delta]^{1/3},
\end{equation}
with the discriminant
\begin{equation} \label{Eq:delta}
\Delta^2\equiv q^3+r^2.
\end{equation}
Then the solutions can be written as
\begin{align} \label{Eq:solutions}
m_1^2&=(s_1+s_2)-\frac{a}{3}, \nonumber \\
m_2^2&=-\frac{1}{2}(s_1+s_2)-\frac{a}{3}+\frac{i\sqrt{3}}{2}(s_1-s_2),
\nonumber \\
m_3^2&=-\frac{1}{2}(s_1+s_2)-\frac{a}{3}-\frac{i\sqrt{3}}{2}(s_1-s_2),
\end{align}
where the $\{m_1,m_2,m_3\}$ refer to the set of masses
$\{M_1,M_2,M_3\}$, but not necessarily ordered.
In the one-loop contribution to the potential (\ref{Eq:V-loop}), only a sum
over $\ell$ enters, the order plays no role.

We recall that the coefficients $a_\alpha=\{a,b,c\}$ that enter in the
cubic eigenvalue equation (\ref{Eq:cubic}) depend on the weak fields
$\eta_i$. In the solutions (\ref{Eq:solutions}), this dependence 
can be accessed via $a$, $s_1$ and $s_2$.
It is thus convenient to write (\ref{Eq:solutions}) more compactly as
\begin{equation}
m_\ell^2=-\frac{a}{3}+\sum_{r=1}^2 A_{\ell r}\,s_r,
\end{equation}
from which it follows that
\begin{equation}
\frac{\partial m_\ell^2}{\partial a_\alpha}
=-\frac{1}{3}\,\delta_{\alpha1}
+A_{\ell r}\frac{\partial s_r}{\partial a_\alpha}.
\end{equation}
Likewise, the higher derivatives are given by
\begin{align}
\frac{\partial^2 m_\ell^2}{\partial a_\alpha\partial a_\beta}
&=A_{\ell r}\frac{\partial^2 s_r}{\partial a_\alpha\partial a_\beta} ,
\nonumber \\
\frac{\partial^3 m_\ell^2}{\partial a_\alpha\partial a_\beta\partial a_\gamma}
&=A_{\ell r}\frac{\partial^3 s_r}
{\partial a_\alpha\partial a_\beta\partial a_\gamma}.
\end{align}

It remains to obtain the derivatives of $s_r$ with respect to $a_\alpha$.  For
this purpose, it is useful to think of $s_r$ as a function of the $q$ and $r$
of Eqs.~(\ref{Eq:q-r})--(\ref{Eq:delta}).  The first derivatives are given by
\begin{equation}
\frac{\partial s_r}{\partial a_\alpha}
=\frac{\partial s_r}{\partial q}\frac{\partial q}{\partial a_\alpha}
+\frac{\partial s_r}{\partial r}\frac{\partial r}{\partial a_\alpha}
=\frac{\partial s_r}{\partial Q^s}\frac{\partial Q^s}{\partial a_\alpha},
\end{equation}
where we collectively refer to $q$ and $r$ as
\begin{equation}
Q^s=\{q,r\}.
\end{equation}

In this notation, the higher derivatives can be writtes as
\begin{equation}
\frac{\partial^2 s_r}{\partial a_\alpha\partial a_\beta}
=\frac{\partial^2 s_r}{\partial Q^s\partial Q^t}
\frac{\partial Q^s}{\partial a_\alpha}\frac{\partial Q^t}{\partial a_\beta}
+\frac{\partial s_r}{\partial Q^s}
\frac{\partial^2 Q^s}{\partial a_\alpha\partial a_\beta},
\end{equation}
and
\begin{align}
\frac{\partial^3 s_r}
{\partial a_\alpha\partial a_\beta\partial a_\gamma}
&=\frac{\partial^3 s_r}{\partial Q^s\partial Q^t\partial Q^u}
\frac{\partial Q^s}{\partial a_\alpha}\frac{\partial Q^t}{\partial a_\beta}
\frac{\partial Q^u}{\partial a_\gamma} \nonumber \\
&+\frac{\partial^2 s_r}{\partial Q^s\partial Q^t}
\biggl[
 \frac{\partial^2 Q^s}{\partial a_\alpha\partial a_\beta}
 \frac{\partial Q^t}{\partial a_\gamma}
+\frac{\partial^2 Q^s}{\partial a_\beta\partial a_\gamma}
 \frac{\partial Q^t}{\partial a_\alpha} \nonumber \\
&+\frac{\partial^2 Q^s}{\partial a_\gamma\partial a_\alpha}
 \frac{\partial Q^t}{\partial a_\beta}
\biggr] 
+\frac{\partial s_r}{\partial Q^s}
\frac{\partial^3 Q^s}
{\partial a_\alpha\partial a_\beta\partial a_\gamma}.
\end{align}
\bigskip

Finally,
the various derivatives $\partial Q^r/\partial a_\alpha$ are given by
\begin{alignat}{2}
\frac{\partial q}{\partial a} &=-\frac{2}{9}a,
&\qquad
\frac{\partial r}{\partial a} &=\frac{1}{6}b-\frac{1}{9}a^2, \nonumber \\
\frac{\partial q}{\partial b} &=\frac{1}{3},
&\qquad
\frac{\partial r}{\partial b} &=\frac{1}{6}a,
\nonumber \\
\frac{\partial q}{\partial c} &=0,
&\qquad
\frac{\partial r}{\partial c} &=-\frac{1}{2},
\end{alignat}

\begin{alignat}{2}
\frac{\partial s_1}{\partial q}
&=\frac{Q^2}{2\Delta\sqrt[2/3]{r+\Delta}},
&\qquad
\frac{\partial s_1}{\partial r}
&=\frac{1+\frac{r}{\Delta}}{3\sqrt[2/3]{r+\Delta}},
\nonumber \\
\frac{\partial s_2}{\partial q}
&=-\frac{Q^2}{2\Delta\sqrt[2/3]{r-\Delta}},
&\qquad
\frac{\partial s_2}{\partial r}
&=\frac{1-\frac{r}{\Delta}}{3\sqrt[2/3]{r-\Delta}}
\end{alignat}

\begin{align}
\frac{\partial s_1}{\partial a}
&=\frac{-a}{9}\frac{Q^2}{\Delta\sqrt[2/3]{r+\Delta}}
+\frac{1+\frac{r}{\Delta}}{3\sqrt[2/3]{r+\Delta}}
\left(\frac{1}{6}b-\frac{1}{9}a^2\right),
\nonumber \\
\frac{\partial s_1}{\partial b}
&= \frac{1}{6}\frac{Q^2}{\Delta\sqrt[2/3]{r+\Delta}}
+\frac{a}{18}\frac{1+\frac{r}{\Delta}}{\sqrt[2/3]{r+\Delta}},
\nonumber \\
\frac{\partial s_1}{\partial c}
&=-\frac{1}{6}\frac{1+\frac{r}{\Delta}}{\sqrt[2/3]{r+\Delta}},
\nonumber \\
\frac{\partial s_2}{\partial a}
&=\frac{a}{9}\frac{Q^2}{\Delta\sqrt[2/3]{r-\Delta}}
+\frac{1-\frac{r}{\Delta}}{3\sqrt[2/3]{r-\Delta}}
\left(\frac{1}{6}b-\frac{1}{9}a^2\right),
\nonumber \\
\frac{\partial s_2}{\partial b}
&=-\frac{1}{6}\frac{Q^2}{\Delta\sqrt[2/3]{r-\Delta}}
+\frac{a}{18}\frac{1-\frac{r}{\Delta}}{\sqrt[2/3]{r-\Delta}},
\nonumber \\
\frac{\partial s_2}{\partial c}
&=-\frac{1}{6}\frac{1-\frac{r}{\Delta}}{\sqrt[2/3]{r-\Delta}}.
\end{align}


\end{document}